\documentclass[journal]{IEEEtran}
\usepackage{cite}
\usepackage{amsmath}
\usepackage[papersize={8.5in,11in}, margin=0.75in]{geometry}
\usepackage{amsfonts}
\usepackage{graphicx}
\usepackage{algorithm}
\usepackage{algpseudocode}
\usepackage{pgfplots}
\pgfplotsset{compat=newest}
\pgfplotsset{plot coordinates/math parser=false}
\usepackage{array}
\usepackage[utf8]{inputenc}



\makeatletter
\newcommand*\titleheader[1]{\gdef\@titleheader{#1}}
\AtBeginDocument{%
  \let\st@red@title\@title%
  \def\@title{%
    \bgroup\normalfont\large\centering\@titleheader\par\egroup
    \vskip0.5em\st@red@title}
}
\makeatother

\title{RESDN: A Novel Metric and Method for Energy Efficient Routing in Software Defined Networks}
\titleheader{Submitted to IEEE Transactions on Network and Service Management}

\begin{document}

\author{\IEEEauthorblockN{Beakal Gizachew Assefa,  Öznur Özkasap, \IEEEmembership{Senior Member, IEEE}}\\
\IEEEauthorblockA{Department of Computer Engineering\\
Koç University, Istanbul, Turkey}\\
\{bassefa13, oozkasap\}@ku.edu.tr}
\maketitle

\IEEEoverridecommandlockouts


\IEEEpubid{\begin{minipage}{\textwidth}\ \\ \ \\ \ \\ \ \\[12pt] 
\\
\\
 \textbf{This work has been submitted to the IEEE for possible publication. Copyright may be transferred without notice, after which this version
may no longer be accessible}
\end{minipage}} 
\begin{abstract}
Software-defined networking (SDN) paradigm, with the flexible and logically centralized control, enables dynamically minimizing the network energy consumption by redirecting paths of packets. However, the links and switches are designed to accommodate maximum traffic volume and their power consumption is not traffic proportional. Moreover, there exists a trade-off between energy efficiency and network performance that need to be considered together. Addressing these issues, we propose an energy efficiency metric named Ratio for Energy Saving in SDN (RESDN) that quantifies energy efficiency based on link utility intervals. We provide integer programming formulation and method for maximizing the RESDN of the network. To the best of our knowledge, RESDN approach is novel as it measures how links are profitably utilized in terms of the amount of energy they consume with respect to their utility. We analyze our approach considering various metrics of interest, and different types of SDN enabled switches. Experiments show that maximizing the RESDN value improves energy efficiency while maintaining acceptable network performance. In comparison to state-of-the-art utility-based heuristics, RESDN method achieves up to 30\% better ratio for energy saving, 14.7 watts per switch power saving, 38\% link saving, 2 hops decrease in average path length, 5\% improved traffic proportionality.

\end{abstract}

\begin{IEEEkeywords}
Software-defined networks, SDN, Energy Efficiency, Ratio for Energy Saving, Routing, Heuristics, Traffic Proportionality. 
\end{IEEEkeywords}

\IEEEpeerreviewmaketitle

\section{Introduction} \label{introduction}

\IEEEPARstart{S}{oftware} Defined Networking (SDN) paradigm is based on the concept of forwarding (data) plane and control plane separation. The control and management of the network is done from a logically centralized controller, while the data plane handles the forwarding functionality. SDN provides the powerful feature of network programmability and has been widely adopted by major companies and network equipment vendors. The ease and flexibility of network control has paved the way to several network applications like load balancing, energy efficiency, dynamic routing, advanced security, and traffic engineering \cite{2019oursurvey,compsurvey,2015scisco,huawei,2018monitorlatency,2019smartgrid,2019engInternet,2018load}.  

Global energy consumption has both environmental and economical issues. 10\% of the global energy consumption is due to ICT sector out of which 2\% is from network components. By 2020 the total electricity cost of cloud data centers is expected to increase by 63\% \cite{stat,MAALOUL2017}. However, most of the time network resources are under utilized as compared to the service they render. 
In data center networks, components are utilized 30\% to 40\% most of the time \cite{dream2019,CarierGrade}. The overall power consumption however remains almost the same for varying amount of traffic volume. A practical solution for a traffic proportional energy consumption problem is to sleep/turn off under-utilized components for low traffic volume. Minimizing the number of network components for low traffic, leads to network performance degradation \cite{ElT,CARPO,2012enablingopenflow,poxsimulation2015,2016joint,orchestrate2016}.

Energy saving at the cost of network performance degradation, however, is an undesirable result. The trade-off between energy saving and performance is quite a challenging task. Energy efficient solutions mainly focus on minimizing the number of energy consuming devices whereas network performance optimization solutions mainly focus on keeping the devices work at full capacity regardless of low traffic. The two opposing objectives  clearly signifies the need for a holistic solution that maintains the trade-off between efficiency and performance.

In SDN, a naive way for measuring the efficiency of a network is to  take the ratio of the power used to the maximum power consumption of the network components. A wide range of energy efficiency metrics like Power Usage Efficiency (PUE) have been proposed and used to measure the efficiency of data centers \cite{greengrid,SPEC}. The Green Grid and Standard Performance Evaluation Corporation (SPEC) are the most widely known efforts among the many. The metrics proposed, however, are either applicable to data centers only or do not consider the utilities of the network resources, and cannot directly be applied to dynamically changing and programmable networks.

Therefore, addressing the need for a metric to measure the energy efficiency of a network with regard to traffic volume and utility of resources, we propose a novel energy efficiency metric, namely RESDN (Ratio for Energy Saving in SDN), that quantifies energy efficiency based on utility intervals defined by minimum and maximum link utility parameters.\footnotemark. \footnotetext{A very preliminary version of this work was presented in \cite{2019EPT}.} 

The contributions of this work are as follows.
\begin{itemize}

    \item We propose a novel energy efficiency metric RESDN: the ratio for energy saving in SDN that quantifies energy efficiency based on link utility intervals. To the best of our knowledge, the approach is unique as it measures how links are profitably utilized in terms of the amount of energy they consume with respect to their utility.
    
	\item We develop Integer Programming (IP) formulation with the objective of maximizing RESDN and propose heuristics algorithm MaxRESDN for achieving the objective.
	
    \item We conduct extensive experiments on Mininet network emulator and POX controller using real network traffic traces and present comparative quantitative analysis of MaxRESDN method. The performance metrics of interest are switch power consumption, RESDN value, percentage of links saved, average path length, throughput, delay, and traffic proportionality.
	
	\item We simulate the power consumption of two hardware switches (NEC and Zodiac FX) and one virtual switch (Open vSwitch-OvS) that are OpenFlow enabled, and conduct comprehensive experiments to measure the power consumption of our proposed MaxRESDN heuristic algorithm.
	
	\item MaxRESDN heuristic algorithm achieves the highest ratio for energy saving which is up to 30\% better than similar state-of-the-art utility-based heuristics algorithms. It also saves up to 38\% links and exhibits an average path length closer to the best energy saving heuristics that focuses on performance. MaxRESDN has the highest traffic proportional energy consumption which 3 to 5\% better than similar utility-based heuristics. To the best of our knowledge, our approach is the first in maintaining the trade-off between energy efficiency, network performance, and traffic proportionality by optimizing RESDN metric. 

	\item Switch power consumption results show that MaxRESDN exhibits on average up to 14.7 watts, 10 watts, and 3.2 watts less power consumption for NEC, OVS and Zodiac FX switches respectively as compared to other utility-based heuristics for energy efficient routing.  
   
    \item As the utility parameters directly impact the performance of the MaxRESDN heuristics, we conduct a detailed analysis of the parameters with respect to a range of traffic volumes.
	\end{itemize}

The remainder of the paper is organized as follows. Section \ref{sec:relatedwork} presents related work on energy efficiency in SDN and energy efficiency metrics. Section \ref{sec:preliminary} presents preliminary work on utility based traffic proportional energy saving in SDN. Section \ref{sec:RESDN } describes RESDN metric, provides IP formulation for RESDN maximization, and presents the MaxRESDN heuristics method that maximizes the RESDN of a network. Section \ref{sec:platform} discusses the platform used for our experiments along with heuristics used for comparison and the characteristics of the hardware switches used. Section \ref{sec:experiment} presents our comprehensive experimental results. Section \ref{sec:conclusion} concludes and states future directions.

\section{Related Work}\label{sec:relatedwork}

The control and forwarding plane separation in SDN has led to flexibility for many network services ranging from load balancing, dynamic routing, energy efficient and flexible network control \cite{sdnsurvey}. The primary challenge of energy efficiency in networking is the fact that energy consumption is not proportional to the volume of traffic. The problem is even more difficult in the traditional networking since there is a very limited flexibility. Traffic-aware energy efficient routing techniques in SDN attempt to make energy consumption proportional to the traffic volume. We discuss related work in traffic-aware energy efficient routing in subsection \ref{sec:trafficaware}. Measuring the energy efficiency of a network environment including data centers have been investigated and various metrics were also proposed. In subsection \ref{sec:metric}, we discuss related work in this area and why we needed a new metrics for SDN.

\subsection{Traffic Aware Energy Efficient Routing Techniques} \label{sec:trafficaware}

Traffic aware energy efficient routing techniques in SDN can be classified based on the energy saving capabilities they focus on and the topology structure they are designed for. The main energy saving capabilities are the links and the switches. Some of the works \cite{ 2012enablingopenflow, CARPO, poxsimulation2015,2016joint,orchestrate2016} focus on links, where some others focus on forwarding switches \cite{response,queueenergy}. There also exist works \cite{ElT,dynamic2014,blackseacomassefa} that consider the combination of links and switches as the energy saving components. Furthermore, queue engineering based techniques consider the arrival of packets and their waiting times to decide per-port power requirement \cite{queueenergy, redel, 2017bandwidth, poxsimulation2015, 2016joint, orchestrate2016,2019eeandquality}.

Based on the topology assumption, energy efficient methods can be classified as general or tailored to a specific network topology. Some methods are general in the sense that they are applicable to any kind of network topology \cite{blackseacomassefa,dynamic2014,2015restorable} and some are only applicable to specific topologies such as Fat-tree, butterfly, and BCube, where Fat-tree and BCube topologies are among the widely used structures used in organizing end systems in data centers \cite{ElT,2012enablingopenflow,sleepandoff,CarierGrade}. 

Table \ref{tbl:classification} presents the summary of traffic aware energy efficient techniques in SDN. The Topology column shows the kind of topology structure the approach is designed for and tested on. The Utility-Based column indicates if the approach focuses on the utility of the links. The Link column shows if the method considers links as energy saving component. Queue Engineering column shows if the approach applies queue engineering techniques.

\begin{table}[ht]
\caption{Classification of Energy Efficient Routing Techniques in SDN}
\label{tbl:classification}
\resizebox{\columnwidth}{!}{
\begin{tabular}{l|llll} 
\textbf{Approach}    & \textbf{Topology}  & \textbf{\begin{tabular}[c]{@{}l@{}}Utility\\ Based\end{tabular}} & \textbf{Link}             & \textbf{\begin{tabular}[c]{@{}l@{}}Queue\\ Engineering\end{tabular}} \\ \hline
ElasticTree \cite{ElT}    & Fat-tree           & -    & \checkmark & -    \\
Carrier Grade \cite{CarierGrade}           & General             & -    & \checkmark & -    \\
CARPO \cite{CARPO}        & Fat-tree           & -    & \checkmark & -    \\
EnableOpenflow \cite{2012enablingopenflow} & Fat-tree           & -    & \checkmark & -    \\
Traffic distribution \cite{garroppo2013does} & General  & \checkmark    & \checkmark & -   \\
GreenSDN \cite{poxsimulation2015}          & Bcubic \& Fat-tree & -    & \checkmark & \checkmark   \\
Fine-grained \cite{bolla2015fine}          & General & \checkmark    & \checkmark & -   \\
OpenNaas\cite{2016joint}  & -  & -    & \checkmark & \checkmark   \\
Orchestrate \cite{orchestrate2016}         & -  & -    & \checkmark & \checkmark   \\
Utility-based \cite{blackseacomassefa}        & General              & \checkmark   & \checkmark & -    \\
Dynamic TA \cite{dynamic2014}              & General               & -    & \checkmark & -    \\
REsPoNse \cite{response}  & -  & -    & -        & -    \\
NetFPGA QE\cite{queueenergy}               & -  & -    & -        & \checkmark   \\
GreenRE \cite{redel}      & General             & -    & \checkmark & -    \\
Bandwidth-aware \cite{2017bandwidth}       & Bcubic \& Fat-tree & - \checkmark   & -        & \checkmark   \\
RA-TAH \cite{sleepandoff}  & Fat-tree           & -    & \checkmark & \checkmark   \\
TE based \cite{newTEmethod14}              & General             & -    & \checkmark & -    \\
Re-Routing\cite{casis2014}                 & General             & -    & -        & \checkmark   \\
Resource-aware \cite{2016rsourcetraffic}   & Fat-tree           & -    & \checkmark & -    \\
SDN /Ethernet \cite{2018energyaware}       & General             & -    & \checkmark & -    \\\hline                 
\end{tabular}
}
\end{table}

Energy efficient techniques use a method of sleeping or turning of unused switches or links when the traffic volume is low and turn them on when traffic volume increases \cite{2012enablingopenflow,2012enablingopenflow,2015resourcemangement,2017bandwidth}. The objective of energy efficiency is reducing the number of active network components. However, this has an adverse effect on the performance of the network hence degrades the network performance. There is a trade-off between the two opposite objectives which are minimizing energy consumption and increasing performance. 

A closely related work considers the remaining bandwidth of links \cite{2017bandwidth}. After formulating the problem using IP, they propose a scheduling algorithm. The heuristics proposed schedules paths for the new flows by considering the paths followed by the preceding flows. Unlike this approach, our focus is not on the remaining bandwidth of the links but the utilities of the links. We measure how much each link is utilized based on the utility interval defined by the minimum and maximum utility parameters.

In contrast to the approach of turning off links for energy saving \cite{CARPO,puelimit1,redel,blackseacomassefa}, the work \cite{garroppo2013does} examines if turning off / sleeping switches always result in energy saving. The network is modeled as a graph with the nodes as the switches and the edges as the links. Comprehensive experiments show for cases where the routing power consumption is cubic with respect to traffic volume, distributing the traffic to underutilized links exhibit a better energy saving as compared to turning them off. Since the routing power consumption is polynomial in general, distributing traffic over underutilized links is only applicable to specific cases \cite{garroppo2013does}.

The other approach close to our work is \cite{dynamic2014} which formulates the energy efficiency problem with mixed integer programming (MIP), then proposes heuristic algorithms. There are four variations of the heuristics where two of which namely Shortest Path First (SPF) and Shortest Path Last (SPL) sort the flows in the order of the shortest path. The other two heuristics namely Smallest Demand First (SDF) and Highest Demand First (HDF) sorts the flows according to the rate flow.

In contrast, our approach is flow order independent, it focuses not only on energy efficiency but also in performance, considers the utility of each link, and proposes a link utility-based energy efficiency metric that simultaneously quantifies efficiency and performance. 

\subsection{Energy Efficiency Metrics }\label{sec:metric}

Data center energy efficiency metrics  have been studied widely with companies and standardization organizations. Some of these are the Green Grid, Standard Performance Evaluation Corporation (SPEC), and Transaction Processing Performance Council (TPC) \cite{2017metrics,2018metricsoverview}. 
The Green grid consortium  proposes several metrics to measure  infrastructure power consumption, carbon dioxide emission, water usage efficiency,and the rate of useful information processed related to resources used. Power Usage Effectiveness (PUE), Data Center Infrastructure Efficiency (DCIE), Carbon Usage Effectiveness (CUE), Water Usage Effectiveness (WUE), and Data Center Productivity (DCP), all of which measure energy efficiency and sustainability.  

Power Usage Effectiveness (PUE), the most prevalent metric, is the ratio of the total annual amount of energy that comes into a data center to the energy that is used by IT equipment \cite{pue2011}. IT equipment includes workstations, storage, monitors, and network devices. Total data center energy consists of IT equipment energy consumption plus energy for cooling system components, UPS, switch gear, and data center lightening. The ideal value for PUE is 1 which means all the energy in the data center is used by the IT equipment. The main limitations of PUE metric is that it is only applicable to a single building that supports a data center \cite{puelimit1}. The other limitation of PUE is that it does not focus on specifically on links and switches which are the main energy saving capabilities in SDN. 

Data Center Infrastructure Efficiency (DCIE), which is the inverse of PUE, is calculated as the ratio of IT equipment power consumption to total facility power consumption. Carbon Usage Effectiveness (CUE) is the product of the amount of carbon dioxide emitted per kilowatt hour (CEF) and the data center's annual PUE \cite{CUE2010}. 
Water Usage Effectiveness (WUE) is the ratio of the annual site water usage in liters to the IT equipment energy usage in kilowatt hours (Kwh) \cite{WUE2011}. WUE measures the water consumption in relation to IT equipment energy consumption. However, it is only applicable to a single data center site and does not consider network devices which are the main focus of energy saving in SDN. 
Data Center Productivity (DCP) measures the quantity of useful information processing completed relative to the amount of some resource consumed in producing the work \cite{greengrid}. DCP treats the data center as a black box where an power enters in to the data center do some useful task and leaves. An important concern in this is how to quantify a useful work. 

To the best of our knowledge, there is no metric that is fully focused on network components such as switches and links which are energy saving capabilities in SDN. Using the existing metrics in an SDN environment falls short of capturing the utilities of links which is an important energy saving capability of a network environment. 
The other drawbacks in the metrics are that they are measured annually and offline. It makes the necessity for new metrics that can measure dynamically changing network environment an indisputable argument. Failing to do so undermines the flexible network management capability that could have been exploited from SDN.

\section{Preliminaries}\label{sec:preliminary}

Next Shortest Path (NSP) and Next Maximum Utility (NMU) heuristics were proposed in our prior work \cite{blackseacomassefa} that are generally applicable to any network topology and are flow order independent. NSP and NMU first select those links with utility less than $Umin$ as the candidate links, then aim at redirecting flows passing through the candidate links to the next shortest path or the path in the direction of the link with maximum utility, respectively. The rationale behind selecting links with utilities less than $Umin$ as a candidate is to redirect flows that pass through these links to make them inactive. NSP and NMU attempt to put underutilized links to inactive state. By doing so not only does the number of active links is reduced, but also overall utilities of the links increase. Next, we discuss the NSP algorithm in detail as the preliminary, since it attempts to make every link to be utilized above the minimum utility $Umin$ or to be made inactive.

NSP (Algorithm \ref{alg:minpath}) identifies under-utilized links and re-routes traffic flows passing through them to the next shortest alternative path. The algorithm takes topology as a graph $\mathbb{G}$, set of flows $\mathbb{F}$, values $Umin$, and $Umax$ as inputs. It first initializes the link utilities by taking the current utilities of the links, and then selects the under-utilized links and sets the $CandidateList$. Line \ref{NSP:shortest} returns the list of paths from $Z_{i}$ to $Z_{j}$  excluding those links in the $CandidateList$. The paths stored in $shortestpaths_{ij}$ are sorted according to the path length. The $shortestpaths_{ij}$ is set to $Path_{ij}$ (line \ref{SPN:shortpathes}). NSP algorithm then picks one of the shortest paths $SP$ to replace the direct link. $SP$ is replaced with the next $Path_{ij}$ until  all the links in $\forall e_{ab}\in SP$, $U_{ab}+U_{ij} \leq Umax$ (lines \ref{NSP:Umaxcheck} to \ref{NSP:Umaxcheckend}). Replacing a direct link with alternative path needs redirecting the flows passing through it. Basically, the utilities of the hops composing $SP$ are increased and the utility of the direct link are decreased (lines \ref{NSP:increment} and\ref{NSP:decrement}). The status of a link is updated if the utility is 0 (all the flows passing are redirected to alternative path $SP$), hence updates the graph $\mathbb{G}$'s status (lines \ref{NSP:upstart}-\ref{NSP:apend}). Algorithm \ref{alg:minpath} returns the updated sub graph of $\mathbb{G}$.

\begin{algorithm}[ht]
	\caption{NSP: NextShortestPath($\mathbb{G}$, $\mathbb{F}$, $Umin$,$Umax$)}\label{alg:minpath}
	\begin{algorithmic}[1]    
	\State \textbf{Input: } Graph $\mathbb{G}$, set of traffic flow  $\mathbb{F}$, minimum utility  $Umin$, and maximum utility $Umax$
	\State \textbf{Output: } Sub graph of $\mathbb{G}$
		\State $\mathbb{U} \gets  NetworkStatus(\mathbb{G},\mathbb{F})$  
		\State $CandidateList  =  \label{NSP:candidate} ReturnCandidateList(\mathbb{U},Umin)$   \label{NMU:candidate}
		\ForAll{\texttt{$e_{ij} \in CandidateList$}}
		\State $Path_{ij} \gets  shortestpaths_{ij}$ \label{NSP:shortest}
		\State $SP  =  Path_{ij}[0] $ \Comment{Pick one of the shortest paths} \label{SPN:shortpathes}
		\While{$\exists \text{ } e_{ab} \in SP$ where $U_{ab} +U_{ij} > Umax$} \label{NSP:Umaxcheck}
		\State \texttt{$SP =  \text{next}(Path_{ij})$} 
		\EndWhile \label{NSP:Umaxcheckend}  
		
		\ForAll{$f \texttt{ passing through } e_{ij}$}
		\ForAll{$e_{ab} \in SP$}
		\State \texttt{$U_{ab} =  U_{ab}+\dfrac{\lambda_{f}}{W_{ab}}$} \Comment{increment $U_{ab}$\label{NSP:increment}}
		\EndFor
		\State \texttt{$U_{ij} =  U_{ij}-\dfrac{\lambda_{f}}{W_{ij}}$} \Comment{decrement $U_{ij}$} \label{NSP:decrement}
		\EndFor
		\If{$U_{ij}==0$} \label{NSP:upstart}
		\State $L_{ij} =  0$ \Comment {state of link is inactive} \State \texttt{\text{Turn off}($e_{ij}$} 
		\State \texttt{Update G}
		\EndIf  \label{NSP:apend}         
		\EndFor
	\end{algorithmic}
\end{algorithm}


\section{RESDN Metric: Ratio for Energy Saving in SDN}\label{sec:RESDN }

In this section, we propose RESDN energy efficiency metric based on link utility interval defined by the minimum and maximum utility parameters $Umin$ and $Umax$. Then, we present IP formulation and heuristics to maximize the RESDN value of a network in a dynamic environment.

\subsection{Motivation and Definition}
Traffic aware energy saving approaches attempt to turn off/sleep network components by observing the traffic volume. Traffic proportional energy consumption is a more stricter requirement than traffic-awareness. It demands the percentage of power consumption of network components to be proportional to the traffic volume. However, none of the existing energy efficiency metrics discussed in Section \ref{sec:relatedwork} take the utilities of the network components under consideration with regard to the network traffic volume. 
Addressing this issue, we propose Ratio for Energy Saving in SDN (RESDN) metric that indicates the percentage of links with utilities within the interval defined by $Umin$ and $Umax$ with respect to the total number of active links. 

The RESDN value reflects how the network components are profitably utilized. The current mode of payment for customers in cloud computing is Pay-as-you-go (PAYG) which charges based on the usage of computing resources. Similarly, RESDN enables the network provider to implement Pay-as-you-use (PAYU) in SDN for energy consumption. A simple scenario, in this case, is a network provider aiming to pay the energy cost of a network component only if it is utilized within a utility interval expressed in terms of RESDN. This RESDN policy is expressed by $Umin$ and $Umax$  parameters and are enforced by the controller. The controller decides the links and their corresponding utilities to match the utility interval specified by the provider. The technical solution to this is to redirect the flows passing through underutilized links to other links. Especially in a dynamic environment where the traffic volume increases and decreases, the RESDN value changes, the controller is responsible for keeping the RESDN value at maximum.  With this, the energy cost of the network provider would be proportional to the traffic volume being streamed in the network at any time.

The RESDN metric enables to measure the energy efficiency of the network by calculating the ratio of the number of links with utilities between the $Umin$ and $Umax$ to the total number of active links. Thus, it allows the network provider to proactively set the RESDN value.

Table \ref{tbl:symbol} shows the parameters used in the formulation of the RESDN . The network is modeled as a weighted graph  $\mathbb{G}$= ($\mathbb{Z}$,$\mathbb{E}$) with  $\mathbb{Z}$ as the set of switches where $Z_{i} \in \mathbb{Z}$ represents switch $i$ and $e_{ij}\in \mathbb{E}$ represents that there exists a link between switches $Z_{i}$ and $Z_{j}$. The weight $W_{ij}$ corresponds to the bandwidth of  the link $e_{ij}$. 

\begin{table*}[ht]
\centering
\caption{Table of Symbols}
\label{tbl:symbol}
\resizebox{1.6\columnwidth}{!}{
\begin{tabular}{l|l}
\hline 
\textbf{Parameters} & \textbf{Description}  \\ \hline 
$\mathbb{G}$     & A graph that consists of $\mathbb{Z}$ and $\mathbb{E}$\\ 
$\mathbb{Z}$& Set of switches $Z_{i} \in \mathbb{Z}$ represents switch $i$     \\ 
$S_{i}$     & Binary variable of switch $Z_{i}$       \\
$\mathbb{E}$& Set of links $e_{ij} \in \mathbb{E}$ is a link between switches $Z_{i}$ and $Z_{j}$ \\ 
$L_{ij}$& Binary variable of link $e_{ij}$ (active or inactive)    \\ 
$W_{ij}$    & Bandwidth of the link e$_{ij}$  \\ 
$\mathbb{F}$ & Set of flows $f \in\mathbb{F}=(sr,ds,\lambda _{f}$)   \\ 
$f$ & A flow f=$(sr,ds,\lambda _{f})$ with  source , destination and flow rate    \\ 
$F_{ij}$    & Binary variable of flow $f$ passing through link $e_{ij}$      \\ 
$\mathbb{U}$& Set of utilities of all links\\ 
$U_{ij}$    & Utility of  link $e_{ij}$     \\ 
$X_{ij}$    & Binary variable of link $e_{ij}$ $Umin \leq U_{ij} \leq Umax$ \\ \hline \hline
\end{tabular}
}
\end{table*}

Let binary variable $S_{i}$ denote the status of switch $Z_{i}$ such that
\begin{align*}
& S_{i}=\left\{
\begin{array}{@{}lll@{}}
1, & \text{if switch}\  Z_{i} \text{ is\ active }\\
0, & \text{otherwise}\
\end{array}\right.
\end{align*}

Traffic in the network is represented by the set of flows  $\mathbb{F}$ where $f \in \mathbb{F}$ is defined as $f$=($sr,ds,\lambda _{f}$). $sr$ and $ds \in \mathbb{Z}$ are the source and destination switches of flow $f$, and $\lambda _{f}$ is the rate of $f$ measured in bits per second. 

\begin{align*}
& F_{ij}=\left\{
\begin{array}{@{}lll@{}}
1, & \text{if flow }\ f \text{ passes through }\text{edge } e_{ij} \\
0, & \text{otherwise}\
\end{array}\right.
\end{align*} 

$\mathbb{U}$ is the set of the utilities of every edge in the graph  $\mathbb{G}$ where  $U_{ij}\in \mathbb{U}$

\begin{align}
U_{ij}=\dfrac{\sum_{\forall f}F_{ij} x \lambda _{f}}{W_{ij}}
\end{align}

is defined as the ratio of the sum of the rates of the flows passing through the edge $e_{ij}$ to the link bandwidth $W_{ij}$. Utility of a link is between 0 and 1, where 0 means no flow is passing through the link and 1 means the sum of the flow rates passing through the link is equal to the link bandwidth. Let $Umin$ be the minimum utility value to keep a link active and $Umax$ is the maximum utility of a link.

\begin{align*}
& X_{ij}=\left\{
\begin{array}{@{}lll@{}}
1, & Umin \leq U_{ij} \leq Umax \\
0, & \text{otherwise}\
\end{array}\right.
\end{align*} 

The energy efficiency metric  is formally defined as

\begin{align}
    RESDN = \dfrac{\sum_{\forall e_{ij}} X_{ij}}{\sum_{\forall e_{ij}} L_{ij}}  \label{eq:RESDN }
\end{align}
where $L_{ij}$ is binary variable  that denotes the status of edge $e_{ij}$
\textcolor{black}{\begin{align*}
& L_{ij}=\left\{
\begin{array}{@{}lll@{}}
1, & \text{if edge}\  e_{ij} \text{ is\ active }\\
0, & \text{otherwise}\
\end{array}\right.
\end{align*} }

According to Equation \ref{eq:RESDN }, RESDN value of 1 means that all links that are turned on are operating between the interval defined by parameters $Umin$ and $Umax$. RESDN value of 0 means that none of the links are operating profitably which means that the network is underutilized. 
The motivation behind the RESDN comes from the idea that network providers want to pay for energy consumption for a given resource only if it is utilized at least to a minimum value of $Umin$. If the utility of a resource is less than $Umin$, then it is underutilized. $Umax$ needs to be less than 1 since setting it to 1 would create network congestion which in turn degrades performance. The larger the value of RESDN , the more the percentage of the active links that operate between the $Umin$ and $Umax$ values. Smaller RESDN value shows that a large number of active links operating unprofitably.

\subsection{IP formulation for RESDN Optimization}

The objective function of our IP model is to maximize the RESDN value of the network.

\begin{align}
\text{maximize } & RESDN  \label{eq:trObjective}\\ 
\text{subject to }& \sum_{\forall f}F_{ij}x\lambda _{f} \leq W_{ij}   \text{ , }\forall e_{ij}\label{eq:utcapacity1}\\  
& \sum_{\forall f}F_{ai}=\sum_{\forall f}F_{ib} \textit{ } , \text{ } Z_{i} \neq f^{sr}, Z_{i} \neq f^{ds} \label{eq:utflow1}\\ 
&  F_{mj}=F_{in}  \textit{  , }\forall f \textit{ } Z_{m}=f^{sr} \textit{ , } Z_{n}=f^{ds} \textit{ , } \forall e_{mj} \textit{ , } \exists e_{in}\label{eq:trflow2}\\ 
 & F_{ij}\leq S_{j} \textit{ }and \textit{ }  F_{ji}\leq S_{j} \text{  , } \forall Z_{j} \in \mathbb{Z} \label{eq:linkswitch}\\ 
 & S_{i} \leq \sum_{\forall f}[F_{ij} + F_{ji} ]\text{ },\text{ } \forall Z_{i} \in \mathbb{Z} \label{eq:flowswitch} \\
 & L_{ij} \leq S_{i}\text{ and  }L_{ij} \leq S_{j} \text{ } \forall Z_{i}, Z_{j}\in \mathbb{Z}   \label{eq:linkswitch4}  
\end{align}

The constraint in Equation \ref{eq:utcapacity1} states that the sum of the rates of flows between two switches should not exceed the link capacity. The constraint in equation \ref{eq:utflow1} states that the number of flows entering and leaving switches which are neither destination nor sources of flow should be equal. The constraint in equation \ref{eq:trflow2} assures a flow entering from source switch should reach the destination switch. The constraint in \ref{eq:linkswitch} asserts that a flow should not be assigned to a link that is connected to an inactive switch. Constraint in equation \ref{eq:flowswitch} models the relationship between the flows passing through a link and a switch. It asserts that if no flow is passing through all the links connected to switch Z$_{i}$, then change the binary variable S$_{i}  =  0$. The constraint in Equation \ref{eq:linkswitch4} asserts that a link should be put in active state if and only if both of the switches it is connecting are active, otherwise, it is inactive. 

\subsection{MaxRESDN Heuristics Method}
The energy efficient routing is an NP-hard problem \cite{ElT,CARPO,2018energyaware}. Formal solutions can best fit for a small number of network switches but fail to scale up when the number is in the order of hundreds. However, in cloud data centers where the number of physical machines is in the order of thousand, the number of switches is in the order of hundreds. That triggers the need for heuristics algorithms that can run in reasonable polynomial time to provide sub-optimal solutions. 

\begin{algorithm}
\caption{MaxRESDN ($\mathbb{G}$, $\mathbb{F}$,$\mathbb{U}$,$Umin$,$Umax$)}\label{alg:MaxRESDN }
	\begin{algorithmic}[1]  		
		\State \textbf{Input: } Graph $\mathbb{G}$, set of traffic flow  $\mathbb{F}$, utility of links $\mathbb{U}$, minimum utility $Umin$, and maximum utility $Umax$
	\State \textbf{Output: } Modified utility of links $\mathbb{U}$ and graph $\mathbb{G}$
		\ForAll{\texttt{$f=(sr,ds,\lambda_{f}) \in F$}} \label{RESDN :flow}
		\State \texttt{$path_{f} \gets$  PathMaxRESDN (sr,ds,$\lambda_{f})$} \label{RESDN :pathmaxi}
		\ForAll{\texttt{$e_{ij} \in path_{f}$}} 		
		\State \texttt{$U_{ij} \gets  U_{ij}+\dfrac{\lambda_{f}}{W_{ij}}$}  \label{RESDN :Uupdate}
		\EndFor  
		\EndFor
		\ForAll{\texttt{$e_{ij} \in \mathbb{E}$}} \label{RESDN :updategraph}		
		\If{$U_{ij} == 0$}
    		    \State $L_{ij} \gets  0$ 
    		   \EndIf
		\EndFor \label{RESDN :updatgraphend}
	\end{algorithmic}
\end{algorithm}

We propose the heuristics algorithm named MaxRESDN (Maximize RESDN). Algorithm \ref{alg:MaxRESDN } aims to maximize the RESDN value of the network as it assigns a path to each flow. The inputs for the MaxRESDN are the network represented as graph $\mathbb{G}$, the current utility of links $\mathbb{U}$, set of flows $\mathbb{F}$, the minimum utility $Umin$ and maximum utility  $Umax$. Each flow $f=(sr,ds,\lambda_{f})\in \mathbb{F}$ is expressed as source $sr$, destination $ds$, and flow rate $\lambda_{f}$ which correspond to source address, destination address, and rate of $f$ measured in bits per second (line \ref{RESDN :flow}). Line \ref{RESDN :Uupdate} increments the utilities of all the links on the path assigned to flow $f$ by the $\dfrac{\lambda_{f}}{W_{ij}}$ factor. Line \ref{RESDN :pathmaxi} assigns  each flow to the path that maximizes the RESDN among alternatives using the $PathMaxRESDN$ (sr,ds,$\lambda_{f})$ method. Lines \ref{RESDN :updategraph} to \ref{RESDN :updatgraphend} makes link $e_{ij}$ inactive if its corresponding utility $U_{ij}$ is 0.

\begin{algorithm}
\caption{PathMaxRESDN(sr,ds,$\lambda _{f}$)}\label{alg:pathmaximizing }%
	\begin{algorithmic}[1]  		
		\State \textbf{Input: } Graph $\mathbb{G}$,utility of links $\mathbb{U}$, minimum utility $Umin$, maximum utility $Umax$, source node sr, destination node ds, and flow rate $\lambda_{f}$
	\State \textbf{Output: } {path$_{f}$ for flow f$(sr,ds,\lambda_{f})$ that maximizes RESDN}
	     \State $AllPath \gets$ all paths between sr and ds \label{maximize:allpath}
		\ForAll{\texttt{Path P $\in AllPath$ }} \label{maximize:pinpath}
		\ForAll {links $e_{ab} \in$  P }\label{maximize:umax}
		    \If{$U_{ab} + \lambda_{f} > UMax$} 
    		    \State  $AllPath$.remove(P) 
    		  \EndIf
    	  \EndFor \label{maximize:umax2}
    	\EndFor
    	\State Path$_{f} \gets  MaxRESDN(Allpath)$ \label{maximize:maxpath}
	\end{algorithmic}
\end{algorithm}

Algorithm \ref{alg:pathmaximizing } (PathMaxRESDN) returns the path that maximizes RESDN for a given flow f with source sr, destination ds, and flow rate $\lambda_{f}$. Line \ref{maximize:allpath} initializes the $Allpath$ list with all paths from sr to ds of flow f. Lines from \ref{maximize:umax} to \ref{maximize:umax2}, ensure the stability of the network by making the utilities of all links in each path plus the rate of flow of f not to exceed the $Umax$ value. While line \ref{maximize:maxpath} selects the path with maximum RESDN value and assign it to  Path$_{f}$.

\section{Experimental Platform}\label{sec:platform}

\textcolor{black}{This section presents the platform used to conduct our experiments. First, the simulation setup and the performance metrics are discussed, then utility based heuristics algorithms used for comparison are presented. Finally, the characteristics of the three OpenFlow protocol enabled switches we used in the experiments for measuring power consumption are presented.} 

\subsection{Setup and Performance Metrics}
The experimental platform is constructed using POX controller and Mininet \cite{mininet} network emulator installed on Ubuntu 16.04 64-bit. The network topology is created on Mininet, and the heuristics are implemented on POX controller. Our experiments are conducted using real traces from SNDlib \cite{SNDlib10}, in particular the GEANT dynamic traffic trace of the European research network.

\begin{figure}[htb!]
\centering
\minipage{0.50\textwidth}
  \includegraphics[width=\linewidth,trim=8 8 8 8,clip]{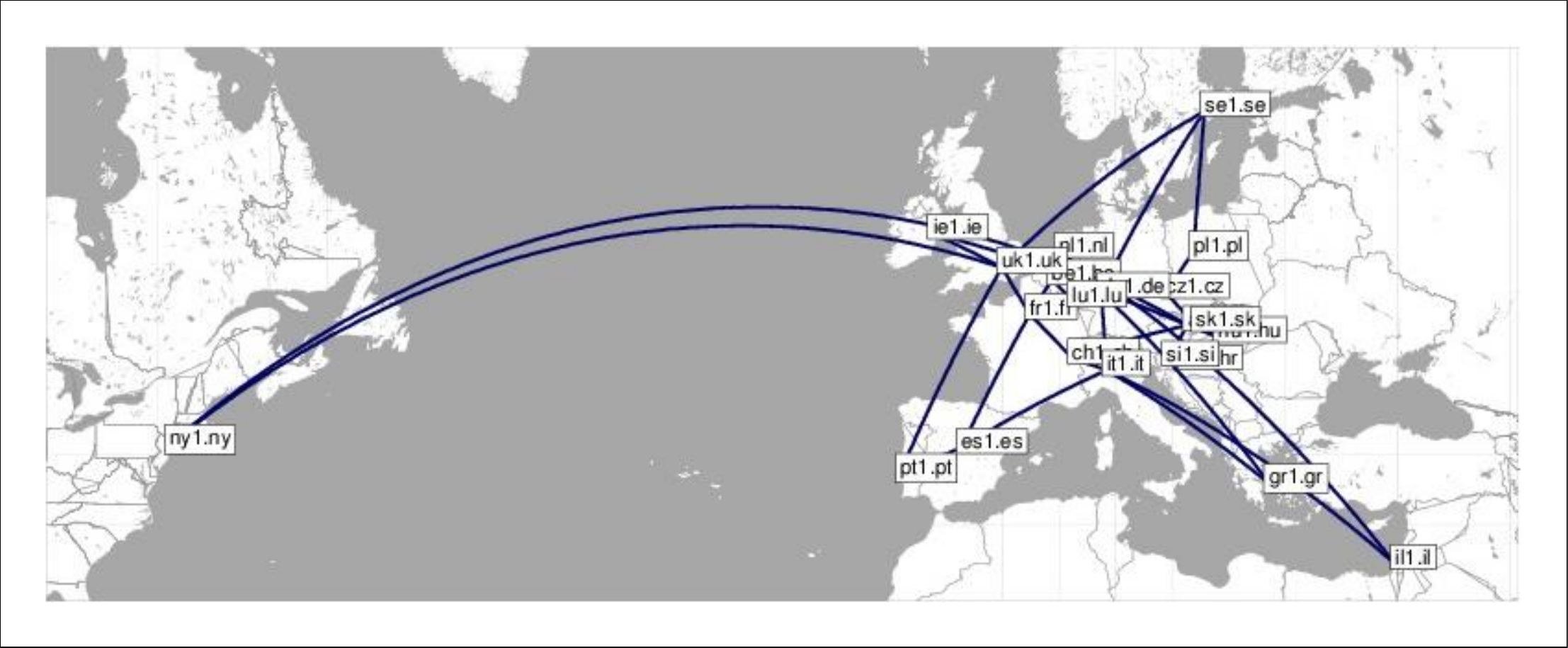}
\endminipage
\hspace{1.706cm}
\minipage{0.45\textwidth}
  \includegraphics[width=\linewidth]{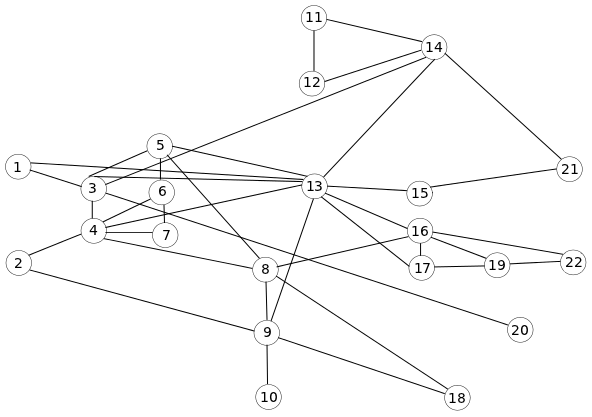}
  
\endminipage 

\caption{GEANT network topology  a) Map view, b) Graph view } \label{fig:GEANT}
\end{figure}

Figure \ref{fig:GEANT} shows the GEANT topology used in our experiments \cite{SNDlib10}. The number of nodes and bidirectional links are 22 and 36, respectively. Traffic demand matrices are aggregated for 15 minutes from a 4-month duration trace. There exists total 11460 traffic demand matrices. The number of flows in each traffic demand matrix ranges from 82 to 462. To show traffic awareness in our experiments, different traffic volumes ranging from 20\% to 90\% with respect to the network capacity are used. 

\begin{table}[htb!]
\centering
\caption{Table of Metrics and Measurement Units}
\label{tbl:metrics}
\resizebox{1\columnwidth}{!}{%
\begin{tabular}{l|l}
\hline
Metric & Unit / Description \\ \hline
Average Power Consumption of Switches         & Watt     \\ 
RESDN          & metric as percentage \\ 
Links saved          & percentage of links saved  \\
Average path length        & \#hops \\
Throughput         & Mbits/sec    \\ 
Delay         & milliseconds    \\
Traffic Proportionality & [0,1] \\ \hline \hline
\end{tabular}
}

\end{table}

Table \ref{tbl:metrics} presents the performance metrics of interest measured in our experiments, namely average power consumption of switches, RESDN value, percentage of links saved, average path length, throughput, delay, and traffic proportionality. The network performance is analyzed in terms of average path length, throughput and delay.

\begin{align}
    P_{switch} = P_{base} + P_{config} + P_{control} 
\label{eq:power}
\end{align}

Equation \ref{eq:power} shows the total power consumption of a switch (P$_{switch}$) as the sum of P$_{base}$, P$_{config}$ and  P$_{control}$ \cite{powermeasure2014}. P$_{base}$ is the power consumption for keeping the switch on without any active ports. {The configuration power consumption P$_{config}$ is calculated as }

\begin{align}
   P_{config}=\sum_{i}^{N_{activePorts}}c_{i}.P_{port} \label{eq:Config}
\end{align}

where {$c_{i}$ is the percentage of the maximum line speed of the port and $P_{port}$ is the power consumption of a port at full capacity measured in watt.}

Equation \ref{eq:control} shows the power consumption of control $P_{control}$ where $r_{PacketIn}$ and $E_{PacketIn}$ are the rate and energy consumption of $PacketIn$. $r_{FlowMod}$ and $E_{FlowMod}$ are the rate and energy consumption of the $FlowMod$ operations.

\begin{align}
    P_{control}=r_{PacketIn}xE_{PacketIn}+r_{FlowMod}xE_{FlowMod}  \label{eq:control}
\end{align}

The ratio of energy saving metric for SDN is calculated as 
\begin{align*}
    RESDN = \dfrac{\sum_{\forall e_{ij}} X_{ij}}{\sum_{\forall e_{ij}} L_{ij}}
\end{align*}
where $L_{ij}$ is binary variable that denotes the status of edge $e_{ij}$ and $X_{ij}$ is a binary variable that tells if the utility of the link is between $Umin$ and $Umax$. Although, our primary goal is energy saving, RESDN also captures performance by the $Umax$ value. 

The percentage of links saved is calculated as 
\begin{align}
\text{Links saved} =  100(1-\dfrac{\sum_{\forall e_{ij}}L_{ij}}{|\mathbb{E}|})  \label{eq:linksaved}
\end{align}

The average path length in terms of average number of hops is calculated as  
\begin{align}
\text{Average path length}  =  \dfrac{\sum_{\forall f}\sum_{\forall e_{ij}}(F_{ij})}{|\mathbb{F}|} \label{eq:avpath}
\end{align}

Throughput is calculated as the amount of data transferred per unit of time. In Mininet, we use Iperf command to measure the throughput of source and destination pairs. In the experiments, we measure the average throughput of all source and destination pairs.

Delay is calculated as the amount of time needed for data packets to be transferred from their source to destination. In our experiments, network delay is measured as the time it takes for the packets of a flow to start at the source and reach at the destination node. The delay is measured both on the Mininet side and POX controller.

Traffic proportionality is defined as the ratio of the traffic volume percentage to the network power consumption and is computed as


\begin{align}
    (\dfrac{\% \text{Traffic Volume}}{M  \dfrac{\sum_{\forall Z_{i}}SP_{i} }{ |\mathbb{Z}|} +  N \dfrac{\sum_{\forall e_{ij}}L_{ij}}{|\mathbb{E}|}} )(M+N) \label{eq:trafficpropotionality}
\end{align}
 where M:N is the ratio of the power consumption of the switches to the links. In these experiments, the ratio M:N is used as 3:1 \cite{blackseacomassefa}.

\subsection{Algorithms Used for Comparison}

The heuristics algorithms we used to compare with MaxRESDN heuristics are listed in Table \ref{tbl:algorithms}. The methodology in \cite{dynamic2014} formulates the problem with MIP and proposes four heuristic algorithms, namely Shortest Path First (SPF), Shortest Path Last (SPL), Smallest Demand First (SDF), and Highest Demand First (HDF). The first flow is assigned its corresponding shortest path, the succeeding flows are assigned paths where the change in energy consumption is minimized. Each heuristics algorithm uses a different criteria to sort the flows and processes them in the corresponding order. The four variations of the algorithms first sort the flows according to the shortest path first, shortest path last, smallest demand first, and highest demand first.

As explained in Section \ref{sec:preliminary}, the objective of NSP algorithm is to re-route flows passing through the under-utilized links to the next shortest path. NMU, on the other hand, chooses the path that has the link with maximum utility. While NSP gives priority to performance, NMU focuses on maximizing the utility of active links. Both NSP and NMU are not only ordering independent but can also be applied on top of other algorithm outputs to improve efficiency. The B heuristics is the application of NSP or NMU algorithms on top of the results of SPF, SPL, SDF, or HDF algorithms. It refers to the best outcome (B) in terms of energy efficiency.

\begin{table}[htb!]
\centering
\caption{Table of heuristics}
\label{tbl:algorithms}
\resizebox{0.90\columnwidth}{!}{%
\begin{tabular}{l|l}
\hline
Abbreviation & Description                       \\ \hline
SPF  \cite{dynamic2014}        & Shortest Path First     \\
SPL   \cite{dynamic2014}       & Shortest Path Last     \\  
SDF  \cite{dynamic2014}         & Smallest Demand First    \\
HDF   \cite{dynamic2014}        & Highest Demand First    \\ 
NSP  \cite{blackseacomassefa}        & Next Shortest Path                \\
NMU  \cite{blackseacomassefa}       & Next Maximum Utility \\ 
B \cite{blackseacomassefa} & Best Combination of NSP (NMU) with others \\ \hline
MaxRESDN          & Maximize RESDN  \\ \hline
\end{tabular}
}
\end{table}

Table \ref{tbl:parameters} shows the values of $Umin$ and $Umax$ parameters used in the experiments for different percentage of traffic volumes. The values that maximize the number of links saved are determined by a greedy-based approach. A detailed analysis of these parameters is presented in Section \ref{subsec:anlys}.

\begin{table}[htb!]
\centering
\caption{Parameters of $Umin$ and $Umax$ Used}
\label{tbl:parameters}
\resizebox{1\columnwidth}{!}{%
\begin{tabular}{l|llllllll}
\hline 
Traffic volume percentage & 20 & 30 & 40 & 50 & 60 & 70 & 80 & 90 \\  \hline
$Umin$    & 31 & 28 & 30 & 25 & 20 & 19 & 15 & 12 \\
$Umax$    & 82 & 85 & 90 & 90 & 92 & 95 & 95 & 95 \\ \hline
\end{tabular}
}
\end{table}

\subsection{Types of Switches}

Table \ref{tbl:powerprofile} shows the parameters used in the power calculation of two hardware switches (NEC PF 5240 and Zodiac FX) and one virtual switch Open vSwitch (OvS) which are OpenFlow enabled. NEC PF 5240 is a hybrid switch technology that adds OpenFlow protocol to the traditional network functionality \cite{nec,powermeasure2014}. Open vSwitch (OvS) is a multilayer virtual switch designed for enabling network automation by supporting various protocols including but not limited to OpenFlow \cite{ovs,powermeasure2014}. Zodiac FX is an OpenFlow switch designed for small scale uses \cite{zodiac}.

\begin{table}[ht]
\caption{Switch Power Consumption parameters}
\resizebox{1\columnwidth}{!}{%
\begin{tabular}{l|l|l|l} \hline
Parameters& NEC PF 5240 \cite{powermeasure2014,nec}      & OvS \cite{powermeasure2014}       & Zodiac FX \cite{zodiac} \\ \hline
Base[W]    & 118.33  & 48.7397   & 15\\
$P_{port}$[W]                  & 0.52  & Na        & 0.15 \\
$E_{PacketIn}$ [$\mu$ W/packet] & 711.30 & 775.53  & 775.53    \\
$E_{FlowMod}$ [$\mu$ W/packet]  & 29.25  & 356.743 & 1455.13  \\ \hline 
\end{tabular}
}
\label{tbl:powerprofile}
\end{table}

\section{Experimental Results}\label{sec:experiment}
\textcolor{black}{This section presents our extensive experimental results in terms of switch power consumption, RESDN and energy efficiency, network performance, and analysis of the link utility interval parameters. }

\subsection{Switch Power Consumption Results}

\begin{figure}[htb!]
	 \begin{minipage}[b]{.70\textwidth}
		\scalebox{0.60}{
%
%
\definecolor{mycolor1}{rgb}{0,1,0}%
\definecolor{mycolor2}{rgb}{0,0,1}%
\definecolor{mycolor3}{rgb}{0,0,0}%
\definecolor{mycolor4}{rgb}{0,0,0.8039}%
\definecolor{mycolor5}{rgb}{0.75294,0.75294,0.75294}%
\definecolor{mycolor6}{rgb}{0.50196,0.50196,0.50196}%
\definecolor{mycolor7}{rgb}{0,0,0}	
\definecolor{mycolor8}{rgb}{1,0,0}	
\begin{tikzpicture}	
\begin{axis}[%
width=4.521in,	
height=3.566in,	
at={(1.322in,0.742in)},	
scale only axis,	
xmin=20,	
xmax=100,	
ymin=120,	
ymax=220,	
ylabel style={font=\color{white!15!black}},	
ylabel={Power (watt)},	
xlabel style={font=\large},	
xlabel={Traffic Volume (\%)},	
ylabel style={font=\large},	
axis background/.style={fill=white},	
axis x line*=bottom,	
axis y line*=left,	
xmajorgrids,	
ymajorgrids,	
legend style={at={(0.682,0.202)}, anchor=south west, legend columns=1, legend cell align=left, align=left, draw=white!15!black}	
]	
\addplot [smooth, color=mycolor1, line width=1pt, mark size=4.5pt, mark=asterisk, mark options={solid, mycolor1}]	
table[row sep=crcr]{%
20	139.916\\
30	146.009\\
40	153.984\\
50	171.932\\
60	179.371\\
70	190.364\\
80	200.966\\
90	213.914\\
};	
\addlegendentry{NSP}	
\addplot [smooth, color=mycolor2, line width=1pt, mark size=6pt, mark=triangle, mark options={solid, mycolor2}]	
table[row sep=crcr]{%
20	134.485\\
30	145.241\\
40	150.348\\
50	168.481\\
60	181.511\\
70	191.204\\
80	202.898\\
90	209.587\\
};	
\addlegendentry{NMU}	
\addplot [smooth, color=mycolor3, line width=1pt, mark size=6pt, mark=square, mark options={solid, mycolor3}]	
table[row sep=crcr]{%
20	138.045\\
30	148.339\\
40	154.649\\
50	173.576\\
60	183.545\\
70	190.816\\
80	201.674\\
90	210.425\\
};	
\addlegendentry{SPF}	
\addplot [smooth, color=mycolor4, line width=1pt, mark size=6pt, mark=diamond, mark options={solid, mycolor4}]	
table[row sep=crcr]{%
20	139.737\\
30	146.599\\
40	154.54\\
50	172.449\\
60	183.637\\
70	192.88\\
80	202.613\\
90	211.407\\
};	
\addlegendentry{SPL}	
\addplot [smooth, color=mycolor5, line width=1pt, mark size=6pt, mark=pentagon, mark options={solid, mycolor5}]	
table[row sep=crcr]{%
20	139.093\\
30	146.804\\
40	155.814\\
50	172.085\\
60	185.803\\
70	190.94\\
80	199.444\\
90	211.672\\
};	
\addlegendentry{SDF}	
\addplot [smooth, color=mycolor6, line width=1pt, mark size=6pt, mark=o, mark options={solid, mycolor6}]	
table[row sep=crcr]{%
20	132.175\\
30	144.036\\
40	147.601\\
50	169.488\\
60	175.918\\
70	186.775\\
80	195.313\\
90	207.665\\
};	
\addlegendentry{HDF}	
\addplot [smooth, color=mycolor7, line width=1pt, mark size=6pt, mark=|, mark options={solid, mycolor7}]	
table[row sep=crcr]{%
20	132.175\\
30	144.036\\
40	147.601\\
50	168.481\\
60	175.918\\
70	186.775\\
80	195.313\\
90	207.665\\
};	
\addlegendentry{B}	
\addplot [smooth, color=mycolor8, line width=1pt, mark size=6pt, mark=otimes, mark options={solid, mycolor8}]	
table[row sep=crcr]{%
20	125.232\\
30	135.738\\
40	139.831\\
50	161.022\\
60	170.822\\
70	182.21\\
80	192.22\\
90	199.832\\
};	
\addlegendentry{MaxRESDN}	
\end{axis}	
\end{tikzpicture}%
}	
	\end{minipage}\qquad
	\hspace{-1cm}	
	\caption{Power Profile for NEC PF 5240 switch in watt vs a range of traffic volume from 20\% to 90\%}
	\label{fig:nec} 
\end{figure}
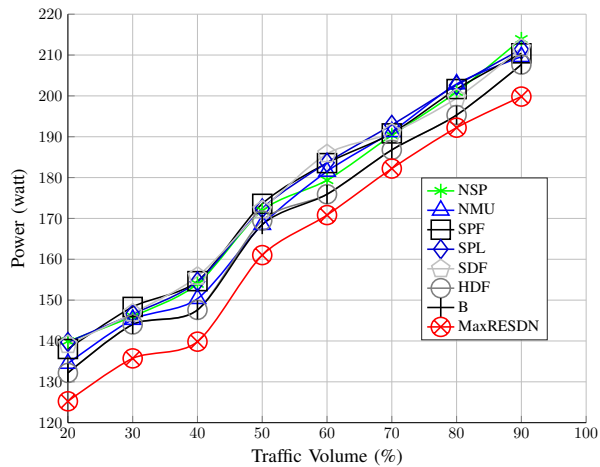

Figure \ref{fig:nec} shows the average power consumption of the switches for the NEC PF 5240 OpenFlow-based switch versus traffic volume. As compared to the best combination (B) of NSP or NMU with other algorithms (Table \ref{tbl:algorithms}), MaxRESDN power consumption is on average 6 to 9 watts less than the B algorithm. As compared to NSP and NMU \cite{blackseacomassefa}, MaxRESDN has shown up to 14.7 and 10.7 watts less in power consumption, respectively. This indicates that through maximizing the RESDN parameter, improvements in energy savings are achieved for all traffic volumes. 

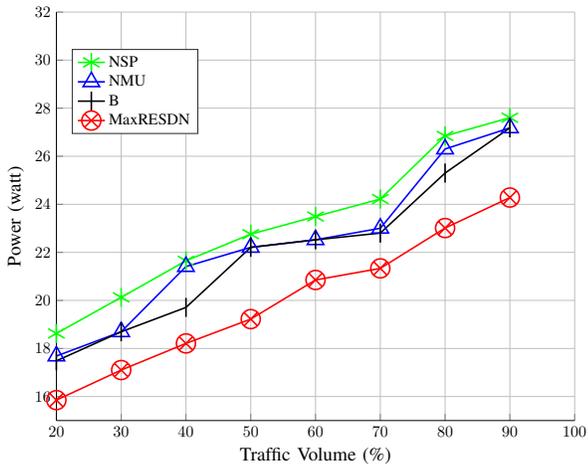
\begin{figure}[htb!]
	 \begin{minipage}[b]{.70\textwidth}
		\scalebox{0.60}{				
\definecolor{mycolor1}{rgb}{0,1,0}%
\definecolor{mycolor2}{rgb}{0,0,1}%
\definecolor{mycolor3}{rgb}{0,0,0}%
\definecolor{mycolor4}{rgb}{0,0,0.8039}%
\definecolor{mycolor5}{rgb}{0.75294,0.75294,0.75294}%
\definecolor{mycolor6}{rgb}{0.50196,0.50196,0.50196}%
\definecolor{mycolor7}{rgb}{0,0,0}											
\definecolor{mycolor8}{rgb}{1,0,0}											
\begin{tikzpicture}											
											
\begin{axis}[%
width=4.521in,											
height=3.566in,											
at={(1.322in,0.742in)},											
scale only axis,											
xmin=20,											
xmax=100,											
ymin=15,											
ymax=32,											
ylabel style={font=\color{white!15!black}},											
ylabel={Power (watt)},											
xlabel style={font=\large},											
xlabel={Traffic Volume (\%)},											
ylabel style={font=\large},											
axis background/.style={fill=white},											
axis x line*=bottom,											
axis y line*=left,											
xmajorgrids,											
ymajorgrids,											
legend style={at={(0.03,0.702)}, anchor=south west, legend columns=1, legend cell align=left, align=left, draw=white!15!black}											
]

\addplot [ color=mycolor1, line width=1.0pt, mark size=6pt, mark=asterisk, mark options={solid, mycolor1}]											
  table[row sep=crcr]{%
20	18.626\\										
30	20.127\\										
40	21.648\\										
50	22.756\\										
60	23.491\\										
70	24.219\\										
80	26.845\\										
90	27.605\\										
};											
\addlegendentry{NSP}											
\addplot [ color=mycolor2, line width=1.0pt, mark size=6pt, mark=triangle, mark options={solid, mycolor2}]											
  table[row sep=crcr]{%
20	17.685\\										
30	18.695\\										
40	21.408\\										
50	22.208\\										
60	22.523\\										
70	23.002\\										
80	26.301\\										
90	27.181\\										
};											
\addlegendentry{NMU}									
											
\addplot [ color=mycolor7, line width=1.0pt, mark size=6.0pt, mark=|, mark options={solid, mycolor7}]											
  table[row sep=crcr]{%
20	17.485\\										
30	18.695\\										
40	19.708\\										
50	22.208\\										
60	22.523\\										
70	22.802\\										
80	25.301\\										
90	27.181\\										
};											
\addlegendentry{B}											
											
\addplot [ color=mycolor8, line width=1.0pt, mark size=6pt, mark=otimes, mark options={solid, mycolor8}]											
  table[row sep=crcr]{%
20	15.85\\										
30	17.102\\										
40	18.21\\										
50	19.222\\										
60	20.851\\										
70	21.333\\										
80	23.012\\										
90	24.281\\										
};											
\addlegendentry{MaxRESDN}

\end{axis}											
\end{tikzpicture}%
}		
	\end{minipage}\qquad
	\hspace{-1cm}	
	\caption{Power Profile for Zodiac FX switch in watt vs a range of traffic volume from 20\% to 90\% }
	\label{fig:zodiac} 
\end{figure}

Figure \ref{fig:zodiac} shows the power consumption of Zodiac FX switch for NSP, NMU, B, and MaxRESDN algorithms versus traffic volume. Similar to the previous experiments, the average power consumption of switches is proportional to the traffic volume. However, power consumption of MaxRESDN is on average 3.2, 2.3, and 2 watts less than NSP, NMU and B algorithms, respectively.

\begin{figure}[htb!]
	 \begin{minipage}[b]{.70\textwidth}
		\scalebox{0.60}{
%
						
\definecolor{mycolor1}{rgb}{0,1,0}%
\definecolor{mycolor2}{rgb}{0,0,1}%
\definecolor{mycolor3}{rgb}{0,0,0}%
\definecolor{mycolor4}{rgb}{0,0,0.8039}%
\definecolor{mycolor5}{rgb}{0.75294,0.75294,0.75294}%
\definecolor{mycolor6}{rgb}{0.50196,0.50196,0.50196}%
\definecolor{mycolor7}{rgb}{0,0,0}											
\definecolor{mycolor8}{rgb}{1,0,0}											
\begin{tikzpicture}											
											
\begin{axis}[%
width=4.521in,											
height=3.566in,											
at={(1.322in,0.742in)},											
scale only axis,											
xmin=20,											
xmax=100,											
ymin=50,											
ymax=130,											
ylabel style={font=\color{white!15!black}},											
ylabel={Power (watt)},											
xlabel style={font=\large},											
xlabel={Traffic Volume (\%)},											
ylabel style={font=\large},											
axis background/.style={fill=white},											
axis x line*=bottom,											
axis y line*=left,											
xmajorgrids,											
ymajorgrids,											
legend style={at={(0.102,0.702)}, anchor=south west, legend columns=1, legend cell align=left, align=left, draw=white!15!black}											
]											
\addplot [ color=mycolor1, line width=1.0pt, mark size=6pt, mark=asterisk, mark options={solid, mycolor1}]											
  table[row sep=crcr]{%
20	62.374\\										
30	67.047\\										
40	76.161\\										
50	82.559\\										
60	92.654\\										
70	100.535\\										
80	110.655\\										
90	122.518\\										
};											
\addlegendentry{NSP}											
\addplot [ color=mycolor2, line width=1.0pt, mark size=6pt, mark=triangle, mark options={solid, mycolor2}]											
  table[row sep=crcr]{%
20	62.844\\										
30	60.099\\										
40	71.033\\										
50	81.558\\										
60	89.52\\										
70	95.035\\										
80	110.103\\										
90	119.46\\										
};											
\addlegendentry{NMU}
\addplot [ color=mycolor7, line width=1.0pt, mark size=6pt, mark=|, mark options={solid, mycolor7}]											
  table[row sep=crcr]{%
20	60.547\\										
30	60.099\\										
40	67.556\\										
50	81.558\\										
60	89.52\\										
70	95.035\\										
80	108.053\\										
90	118.77\\										
};											
\addlegendentry{B}											
											
\addplot [ color=mycolor8, line width=1.0pt, mark size=6pt, mark=otimes, mark options={solid, mycolor8}]											
  table[row sep=crcr]{%
20	52.7843\\										
30	55.893\\										
40	63.22\\										
50	74.34\\										
60	82.23\\										
70	90.08\\										
80	102.3\\										
90	113.783\\										
};											
\addlegendentry{MaxRESDN}

\end{axis}											
\end{tikzpicture}%
}		
	\end{minipage}\qquad
	\hspace{-1cm}	
	\caption{Power Profile for OvS switch in watt vs a range of traffic volume from 20\% to 90\%}
	\label{fig:OvS} 
\end{figure}
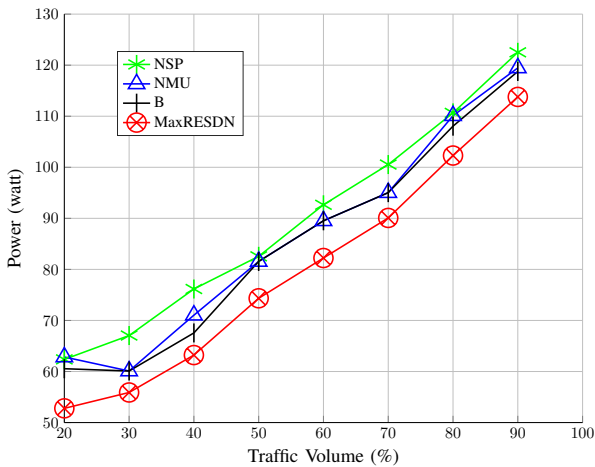

Figure \ref{fig:OvS} shows the power consumption of OvS switches versus traffic volume. Results show that as the traffic volume increases, the average power consumption of switches also increases. The MaxRESDN algorithm exhibits 5.8, 10, and 8.9 watts less energy consuming as compared to B, NSP, and NMU respectively. 

Overall, Figures \ref{fig:nec}, \ref{fig:zodiac}, and \ref{fig:OvS} show that MaxRESDN is the most energy efficient as compared to the other algorithms (Table \ref{tbl:algorithms}). Our findings indicate that maximizing the RESDN value helps minimizing the energy consumption. This is mainly because enforcing links to operate within the minimum and maximum utility parameters minimizes the number of active ports.

\subsection{RESDN and Energy Efficiency Results}

\begin{figure}[htb!]
	\hspace{-.8cm}
	\begin{minipage}[b]{.70\textwidth}
		\scalebox{0.60}{\input{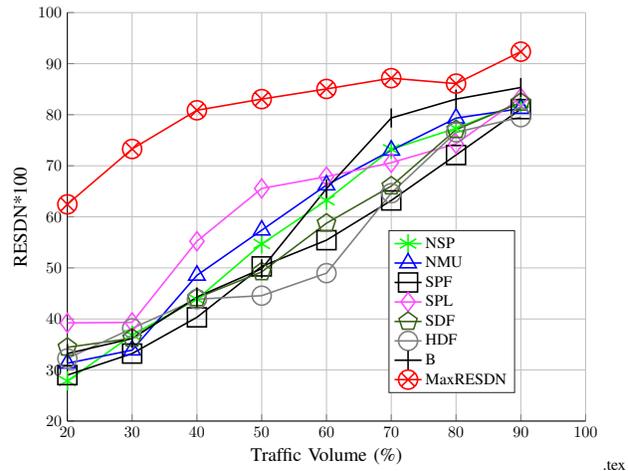}}		
	\end{minipage}\qquad
	\hspace{-1cm}	
	\caption{RESDN Values NSP,NMU, SPF, SPL, SDF, HDF, B and MaxRESDN heuristic algorithms versus a traffic volume ranging from 20\% to 90\%}
	\label{fig:RESDN } 
\end{figure}

Figure \ref{fig:RESDN } shows the RESDN values of all heuristic algorithms as the traffic volume increases. The MaxRESDN algorithm achieves the highest RESDN value which is 22\% better for low traffic and 10\% better than others for high traffic volume. This is because unlike the other algorithms whose objective is to minimize the number of links used in the network, MaxRESDN algorithm aims at maximizing the RESDN value, which increases the utility. In particular, for low traffic, it is more likely for flows that pass through overutilized links to be redirected to underutilized links. 

\begin{figure}[htb!]
	\hspace{2cm}
	
	\begin{minipage}[b]{.70\textwidth}
		\scalebox{0.60}{
%
%
\definecolor{mycolor1}{rgb}{0,1,0}%
\definecolor{mycolor2}{rgb}{0,0,1}%
\definecolor{mycolor3}{rgb}{0,0,0}%
\definecolor{mycolor4}{rgb}{1,0.3,1}%
\definecolor{mycolor5}{rgb}{0.24,0.394,0.10}%
\definecolor{mycolor6}{rgb}{0.50196,0.50196,0.50196}%
\definecolor{mycolor7}{rgb}{0,0,0}	
\definecolor{mycolor8}{rgb}{1,0,0}	
\begin{tikzpicture}	
\begin{axis}[%
width=4.521in,	
height=3.566in,	
at={(1.322in,0.742in)},	
scale only axis,	
xmin=20,	
xmax=100,	
ymin=2,	
ymax=45,	
ylabel style={font=\color{white!15!black}},	
ylabel={Links saved (\%)},	
xlabel style={font=\large},	
xlabel={Traffic Volume (\%)},	
ylabel style={font=\large},	
axis background/.style={fill=white},	
axis x line*=bottom,	
axis y line*=left,	
xmajorgrids,	
ymajorgrids,	
legend style={at={(0.682,0.502)}, anchor=south west, legend columns=1, legend cell align=left, align=left, draw=white!15!black}	
]	
\addplot [ color=mycolor1, line width=1.0pt, mark size=6pt, mark=asterisk, mark options={solid, mycolor1}]	
table[row sep=crcr]{%
20	37.08\\
30	33.90\\
40	25.21\\
50	21.65\\
60	10.13\\
70	8.73\\
80	5.49\\
90	2.75\\
};	
\addlegendentry{NSP}	
\addplot [ color=mycolor2, line width=1.0pt, mark size=6pt, mark=triangle, mark options={solid, mycolor2}]	
table[row sep=crcr]{%
20	37.08\\
30	34.90\\
40	22.21\\
50	15.65\\
60	10.13\\
70	8.23\\
80	5.49\\
90	2.75\\
};	
\addlegendentry{NMU}	
\addplot [ color=mycolor3, line width=1.0pt, mark size=6pt, mark=square, mark options={solid, mycolor3}]	
table[row sep=crcr]{%
20	43.11\\
30	36.10\\
40	32.21\\
50	25.65\\
60	21.3\\
70	10.13\\
80	9.13\\
90  5.45\\	
};	
\addlegendentry{SPF}	
\addplot [ color=mycolor4, line width=1.0pt, mark size=6pt, mark=diamond, mark options={solid, mycolor4}]	
table[row sep=crcr]{%
20	41.89\\
30	37.10\\
40	29.21\\
50	22.36\\
60	20.36\\
70	13.21\\
80	9.98\\
90  5.45\\	
};	
\addlegendentry{SPL}	
\addplot [ color=mycolor5, line width=1.0pt, mark size=6pt, mark=pentagon, mark options={solid, mycolor5}]	
table[row sep=crcr]{%
20	43.11\\
30	37.10\\
40	30.21\\
50	25.65\\
60	21.3\\
70	11.13\\
80	9.13\\
90  2.75\\	
};	
\addlegendentry{SDF}	
\addplot [ color=mycolor6, line width=1.0pt, mark size=6pt, mark=o, mark options={solid, mycolor6}]	
table[row sep=crcr]{%
20	43.89\\
30	37.10\\
40	27.21\\
50	22.36\\
60	15.88\\
70	13.13\\
80	8.5\\
90  6.75\\	
};	
\addlegendentry{HDF}	
\addplot [ color=mycolor7, line width=1.0pt, mark size=6pt, mark=|, mark options={solid, mycolor7}]	
table[row sep=crcr]{%
20	45.22\\
30	38.16\\
40	33.89\\
50	26.84\\
60	26.7\\
70	17.00\\
80	10.05\\
90	7.88\\
};	
\addlegendentry{B}	
\addplot [ color=mycolor8, line width=1.0pt, mark size=6pt, mark=otimes, mark options={solid, mycolor8}]	
table[row sep=crcr]{%
20	38.08\\
30	34.90\\
40	29.21\\
50	25.65\\
60	22.13\\
70	13.21\\
80	9.73\\
90	6.5\\
};	
\addlegendentry{MaxRESDN}	
\end{axis}	
\end{tikzpicture}%
}		
	\end{minipage}\qquad
	\hspace{-1cm}	
	
	\caption{The percentage of links saved versus traffic volume ranging from 20\% to 90\%}
	\label{fig:es} 
\end{figure}
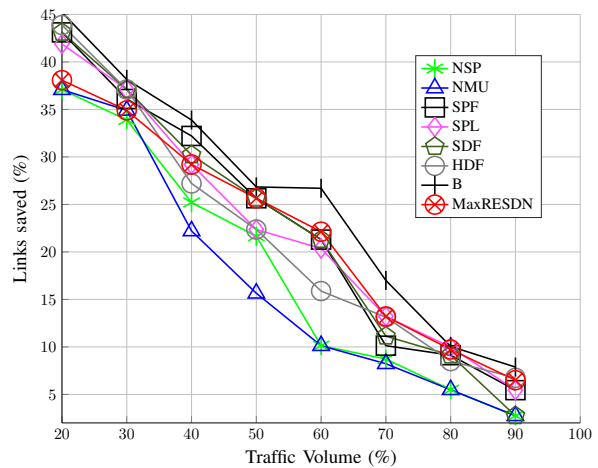

Figure \ref{fig:es} shows the energy saving in terms of the percentage of links put to sleep. The B heuristics which is the best combination of the NSP or NMU algorithm exhibits the highest energy saving, while NSP and NMU have the lowest energy saving. MaxRESDN heuristics saves 38\% for 20\% traffic volume and 6.5\% for 90\% traffic volume. 

\begin{figure}[htb!]
	\hspace{2cm}
	
	\begin{minipage}[b]{.70\textwidth}
		\scalebox{0.60}{\definecolor{mycolor1}{rgb}{0,1,0}%
\definecolor{mycolor2}{rgb}{0,0,1}%
\definecolor{mycolor3}{rgb}{0,0,0}%
\definecolor{mycolor4}{rgb}{0,0,0.8039}%
\definecolor{mycolor5}{rgb}{0.75294,0.75294,0.75294}%
\definecolor{mycolor6}{rgb}{0.50196,0.50196,0.50196}%
\definecolor{mycolor7}{rgb}{0,0,0}											
\definecolor{mycolor8}{rgb}{1,0,0}											
\begin{tikzpicture}											
\begin{axis}[%
width=4.521in,											
height=3.566in,											
at={(1.322in,0.742in)},											
scale only axis,											
xmin=20,											
xmax=100,											
ymin=15,											
ymax=32,											
ylabel style={font=\color{white!15!black}},											
ylabel={Power (watt)},											
xlabel style={font=\large},											
xlabel={Traffic Volume (\%)},											
ylabel style={font=\large},											
axis background/.style={fill=white},											
axis x line*=bottom,											
axis y line*=left,											
xmajorgrids,											
ymajorgrids,											
legend style={at={(0.03,0.602)}, anchor=south west, legend columns=1, legend cell align=left, align=left, draw=white!15!black}											
]

\addplot [ color=mycolor1, line width=1.0pt, mark size=6pt, mark=asterisk, mark options={solid, mycolor1}]											
  table[row sep=crcr]{%
20	18.626\\										
30	20.127\\										
40	21.648\\										
50	22.756\\										
60	23.491\\										
70	24.219\\										
80	26.845\\										
90	27.605\\										
};											
\addlegendentry{NSP}											
\addplot [ color=mycolor2, line width=1.0pt, mark size=6pt, mark=triangle, mark options={solid, mycolor2}]											
  table[row sep=crcr]{%
20	17.685\\										
30	18.695\\										
40	21.408\\										
50	22.208\\										
60	22.523\\										
70	23.002\\										
80	26.301\\										
90	27.181\\										
};											
\addlegendentry{NMU}									
\addplot [ color=mycolor3, line width=1.0pt, mark size=6pt, mark=square, mark options={solid, mycolor3}]											
  table[row sep=crcr]{%
20	19.29\\										
30	20.549\\										
40	21.751\\										
50	23.269\\										
60	23.912\\										
70	24.78\\										
80	27.481\\										
90	28.533\\										
};											
\addlegendentry{SPF}											
\addplot [ color=mycolor4, line width=1.0pt, mark size=6pt, mark=diamond, mark options={solid, mycolor4}]											
table[row sep=crcr]{%
20	18.715\\										
30	20.417\\										
40	22.33\\										
50	22.674\\										
60	22.747\\										
70	24.279\\										
80	27.656\\										
90	28.681\\										
};											
\addlegendentry{SPL}											
\addplot [ color=mycolor5, line width=1.0pt, mark size=6pt, mark=pentagon, mark options={solid, mycolor5}]											
table[row sep=crcr]{%
20	19.929\\										
30	22.935\\										
40	24.51\\										
50	25.216\\										
60	26.216\\										
70	27.976\\										
80	28.434\\										
90	29.012\\										
};											
\addlegendentry{SDF}											
\addplot [ color=mycolor6, line width=1.0pt, mark size=6pt, mark=o, mark options={solid, mycolor6}]											
  table[row sep=crcr]{%
20	17.485\\										
30	18.795\\										
40	19.708\\										
50	22.498\\										
60	22.623\\										
70	22.802\\										
80	25.301\\										
90	27.181\\										
};											
\addlegendentry{HDF}											
											
\addplot [ color=mycolor7, line width=1.0pt, mark size=6.0pt, mark=|, mark options={solid, mycolor7}]											
  table[row sep=crcr]{%
20	17.485\\										
30	18.695\\										
40	19.708\\										
50	22.208\\										
60	22.523\\										
70	22.802\\										
80	25.301\\										
90	27.181\\										
};											
\addlegendentry{B}											
											
\addplot [ color=mycolor8, line width=1.0pt, mark size=6pt, mark=otimes, mark options={solid, mycolor8}]											
  table[row sep=crcr]{%
20	15.85\\										
30	17.102\\										
40	18.21\\										
50	19.222\\										
60	20.851\\										
70	21.333\\										
80	23.012\\										
90	24.281\\										
};											
\addlegendentry{MaxRESDN}						
\end{axis}											
\end{tikzpicture}%
}		
	\end{minipage}\qquad
	\hspace{-1cm}	
	
	\caption{The average power consumption of the Zodiac switch measured in watts versus traffic volume ranging from 20\% to 90\%}
	\label{fig:spower} 
\end{figure}
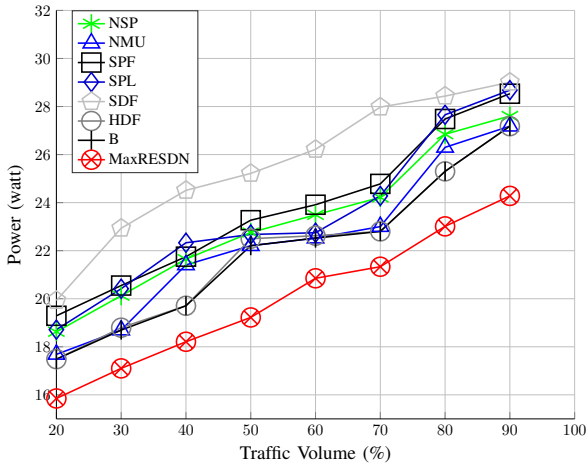

Figure \ref{fig:spower} presents the power consumption of all algorithms for the Zodiac switch. The results show that MaxRESDN switch power consumption is 16 to 22 watts which is on average 4 watts less than the SDF algorithm. MaxRESDN also demonstrated a power consumption on average 2 watts less than the B algorithm. These results show that by maximizing the RESDN value which is based on link utilities, by doing it frees ports on switches, hence it manages to reduce the power consumption of switches. 

\subsection{Network Performance Results}

\begin{figure}[htb!]
		\begin{minipage}[b]{.70\textwidth}
		\scalebox{0.60}{
%
%
\definecolor{mycolor1}{rgb}{0,1,0}%
\definecolor{mycolor2}{rgb}{0,0,1}%
\definecolor{mycolor3}{rgb}{0,0,0}%
\definecolor{mycolor4}{rgb}{0,0,0.8039}%
\definecolor{mycolor5}{rgb}{0.75294,0.75294,0.75294}%
\definecolor{mycolor6}{rgb}{0.50196,0.50196,0.50196}%
\definecolor{mycolor7}{rgb}{0,0,0}											
\definecolor{mycolor8}{rgb}{1,0,0}															
\begin{tikzpicture}

\begin{axis}[%
width=4.521in,
height=3.566in,
at={(1.322in,0.742in)},
scale only axis,
xmin=20,
xmax=100,
ymin=2,
ymax=6,
ylabel style={font=\color{white!15!black}},
ylabel={Average path length (\#hops)},
xlabel style={font=\large},
xlabel={Traffic Volume (\%)},
ylabel style={font=\large},
axis background/.style={fill=white},
axis x line*=bottom,
axis y line*=left,
xmajorgrids,
ymajorgrids,
legend style={at={(0.03,0.45)}, anchor=south west, legend columns=1, legend cell align=left, align=left, draw=white!15!black}
]
\addplot [ color=mycolor1, line width=1pt, mark size=4.5pt, mark=asterisk, mark options={solid, mycolor1}]
  table[row sep=crcr]{%
20	2.7\\
30	2.75\\
40	2.69\\
50	2.99\\
60	3.01\\
70	3.28\\
80	3.48\\
90	3.5\\
};
\addlegendentry{NSP}
\addplot [ color=mycolor2, line width=1pt, mark size=4.5pt, mark=triangle, mark options={solid, mycolor2}]
  table[row sep=crcr]{%
20	2.7\\
30	2.85\\
40	2.99\\
50	3.32\\
60	3.44\\
70	3.48\\
80	3.53\\
90	3.6\\
};
\addlegendentry{NMU}
\addplot [ color=mycolor3, line width=1pt, mark size=4.5pt, mark=triangle, mark options={solid, mycolor3}]
  table[row sep=crcr]{%
20	3.00\\
30	3.22\\
40	3.33\\
50	3.44\\
60	3.55\\
70	3.68\\
80	4.03\\
90	4.5\\
};
\addlegendentry{SPF}
\addplot [ color=mycolor4, line width=1pt, mark size=4.5pt, mark=square, mark options={solid, mycolor4}]
table[row sep=crcr]{%
20	3.20\\
30	3.32\\
40	3.43\\
50	3.44\\
60	3.65\\
70	3.78\\
80	4.23\\
90	4.59\\
};
\addlegendentry{SPL}
\addplot [ color=mycolor5, line width=1pt, mark size=4.5pt, mark=diamond, mark options={solid, mycolor5}]
table[row sep=crcr]{%
20	3.30\\
30	3.34\\
40	3.39\\
50	3.54\\
60	3.75\\
70	3.98\\
80	4.33\\
90	5.4\\
};
\addlegendentry{SDF}
\addplot [ color=mycolor6, line width=1pt, mark size=4.5pt, mark=pentagon, mark options={solid, mycolor6}]
  table[row sep=crcr]{%
20	3.44\\
30	3.58\\
40	3.79\\
50	4.04\\
60	4.45\\
70	4.78\\
80	5.00\\
90	5.6\\
};
\addlegendentry{HDF}

\addplot [ color=mycolor7, line width=1pt, mark size=4.5pt, mark=|, mark options={solid, mycolor7}]
  table[row sep=crcr]{%
20	2.84\\
30	3.18\\
40	3.59\\
50	3.94\\
60	4.25\\
70	4.38\\
80	4.40\\
90	4.54\\
};
\addlegendentry{B}

\addplot [ color=mycolor8, line width=1pt, mark size=4.5pt, mark=otimes, mark options={solid, mycolor8}]
  table[row sep=crcr]{%
20	2.87\\
30	2.95\\
40	3.01\\
50	3.49\\
60	3.61\\
70	3.81\\
80	4.28\\
90	4.65\\
};
\addlegendentry{MaxRESDN}

\end{axis}
\end{tikzpicture}%
}		
	\end{minipage}\qquad
	\hspace{-1cm}	
	\caption{Average path length in terms of number of hops versus traffic volume ranging from 20\% to 90\%}
	\label{fig:aps} 
	
\end{figure}
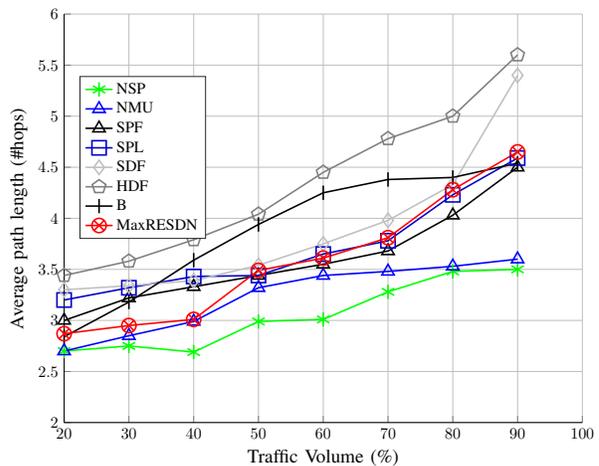

Figure \ref{fig:aps} shows the average path length (in terms of the number of hops) versus the traffic volume, and is calculated as given in equation \ref{eq:avpath}. The NSP heuristics achieves the best path length since its objective is redirecting flows to the next alternative shortest path. The decisions are made periodically by selecting the underutilized links as candidates then redirecting the flows. B heuristics which is the combination of NSP and the best of SPF, SPL, SDF, and HDF in terms of energy saving shows improvement in the average path length. This indicates that our previous heuristics still can be applied on top of other algorithms and improve both performance and energy saving. The MaxRESDN heuristics performance is closer to NMU that has the objective of maximizing the utilities of the links, by directing the flows passing through underutilized links to more utilized links. 

The experimental results show that NSP and NMU have the lowest average path length but least energy saving, since they first consider performance rather than energy efficiency. The other algorithms in \cite{dynamic2014} exhibit better energy saving than NSP and NMU but worst average path length. This is because their objective is energy saving primarily. However, applying NSP and NMU on top of them increases energy saving. NSP and NMU balance the trade-off between performance and energy saving. The MaxRESDN heuristics, as compared to the other heuristics, has the maximum RESDN value. By maximizing a single RESDN value, the MaxRESDN has shown closer results to the best combination algorithm in terms of both energy saving and average path length. Assuming that the optimal value for  $Umin$ and $Umax$ parameters is found for a given traffic, the MaxRESDN heuristics gives the maximum RESDN that achieves traffic proportional energy consumption and keeps the trade-off between energy efficiency and performance. 

\begin{figure}[htb!]
		\begin{minipage}[b]{.70\textwidth}
		\scalebox{0.60}{
%
%
\definecolor{mycolor1}{rgb}{0,1,0}%
\definecolor{mycolor2}{rgb}{0,0,1}%
\definecolor{mycolor3}{rgb}{0,0,0}%
\definecolor{mycolor4}{rgb}{1,0.3,1}%
\definecolor{mycolor5}{rgb}{0.24,0.394,0.10}%
\definecolor{mycolor6}{rgb}{0.50196,0.50196,0.50196}%
\definecolor{mycolor7}{rgb}{0,0,0}											
\definecolor{mycolor8}{rgb}{1,0,0}	
\begin{tikzpicture}	
\begin{axis}[%
width=4.521in,	
height=3.566in,	
at={(1.322in,0.742in)},	
scale only axis,	
xmin=20,	
xmax=100,	
ymin=30,	
ymax=80,	
ylabel style={font=\color{white!15!black}},	
ylabel={Throughput (Mbps)},	
xlabel style={font=\large},	
xlabel={Traffic Volume (\%)},	
ylabel style={font=\large},	
axis background/.style={fill=white},	
axis x line*=bottom,	
axis y line*=left,	
xmajorgrids,	
ymajorgrids,	
legend style={at={(0.02,0.15)}, anchor=south west, legend columns=1, legend cell align=left, align=left, draw=white!15!black}	
]	
\addplot [smooth, color=mycolor1, line width=1.0pt, mark size=6pt, mark=asterisk, mark options={solid, mycolor1}]	
  table[row sep=crcr]{%
20	75.25\\
30	74.53\\
40	75.4\\
50	71.05\\
60	70.76\\
70	66.84\\
80	63.94\\
90	63.65\\
};	
\addlegendentry{NSP}	
\addplot [smooth, color=mycolor2, line width=1.0pt, mark size=6pt, mark=triangle, mark options={solid, mycolor2}]	
  table[row sep=crcr]{%
20	75.25\\
30	73.08\\
40	71.05\\
50	66.26\\
60	64.52\\
70	63.94\\
80	63.22\\
90	62.2\\
};	
\addlegendentry{NMU}	
\addplot [smooth, color=mycolor3, line width=1.0pt, mark size=6pt, mark=square, mark options={solid, mycolor3}]	
  table[row sep=crcr]{%
20	70.9\\
30	67.71\\
40	66.12\\
50	64.52\\
60	62.93\\
70	61.04\\
80	55.97\\
90	49.15\\
};	
\addlegendentry{SPF}	
\addplot [smooth, color=mycolor4, line width=1.0pt, mark size=6pt, mark=diamond, mark options={solid, mycolor4}]	
table[row sep=crcr]{%
20	68\\
30	66.26\\
40	64.67\\
50	64.52\\
60	61.48\\
70	59.59\\
80	53.07\\
90	47.85\\
};	
\addlegendentry{SPL}	
\addplot [smooth, color=mycolor5, line width=1.0pt, mark size=6pt, mark=pentagon, mark options={solid, mycolor5}]	
table[row sep=crcr]{%
20	66.55\\
30	65.97\\
40	65.25\\
50	63.07\\
60	60.03\\
70	56.69\\
80	51.62\\
90	41.32\\
};	
\addlegendentry{SDF}	
\addplot [smooth, color=mycolor6, line width=1.0pt, mark size=6pt, mark=o, mark options={solid, mycolor6}]	
  table[row sep=crcr]{%
20	64.52\\
30	62.49\\
40	62.45\\
50	58.82\\
60	56.88\\
70	55.09\\
80	52.9\\
90	48\\
};	
\addlegendentry{HDF}	
\addplot [smooth, color=mycolor7, line width=1.0pt, mark size=6pt, mark=|, mark options={solid, mycolor7}]	
  table[row sep=crcr]{%
20	63.5\\
30	60.51\\
40	61.12\\
50	57.72\\
60	56.77\\
70	54.87\\
80	51.36\\
90	47.93\\
};	
\addlegendentry{B}	
\addplot [smooth, color=mycolor8, line width=1.0pt, mark size=6pt, mark=otimes, mark options={solid, mycolor8}]	
  table[row sep=crcr]{%
20	73.84\\
30	71.8\\
40	70.91\\
50	68.26\\
60	69.02\\
70	61.8\\
80	58.35\\
90	59.12\\
};	
\addlegendentry{MaxRESDN}	
\end{axis}	
\end{tikzpicture}%
}		
	\end{minipage}\qquad
	\hspace{-1cm}	
	\caption{Throughput of heuristic algorithms in Mbps versus a range of traffic volume from 20\% to 90\%}
	\label{fig:throughput} 
\end{figure}
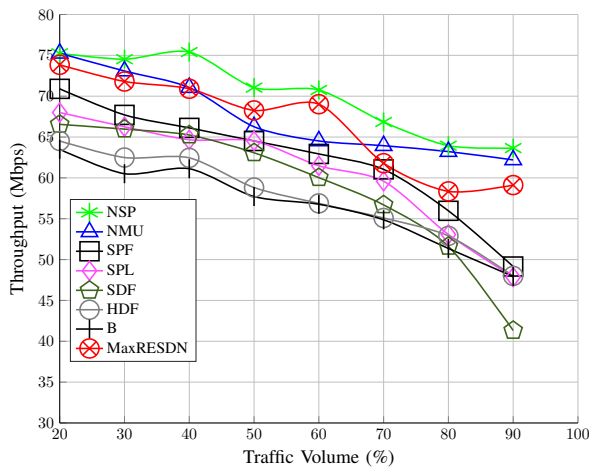

Figure \ref{fig:throughput} shows the throughput of the algorithms in Mbps with respect to the volume of traffic that ranges from 20\% to 90\%. The throughput measurement is done on Mininet using the iperf command. The average bandwidth of links is set to 100 Mbps and the average flow rate is 7.79 Mbps.  Results show that the NSP algorithm exhibits the highest throughput for all the traffic range and is followed by NMU and MaxRESDN. This is because, for the three heuristics, the algorithms are initialized with the shortest paths. The maximum RESDN value by MaxRESDN as shown in figure \ref{fig:RESDN } for all the traffic volume exhibits the least power consumption and also throughput close to that of NSP and NMU. For all the algorithms, the throughput decreases slightly as the traffic size increases after 30\%.  Under normal conditions where energy saving is not applied, throughput is meant to increase until the traffic volume is near 100\% then decreases because of congestion. However, in energy saving routing algorithms, the attempt is to minimize the number of active links and switches. Therefore, the average throughput would decrease even when the traffic volume is near 50\%. The trend of throughput in energy saving algorithms is different from non-energy saving performance focused heuristics.

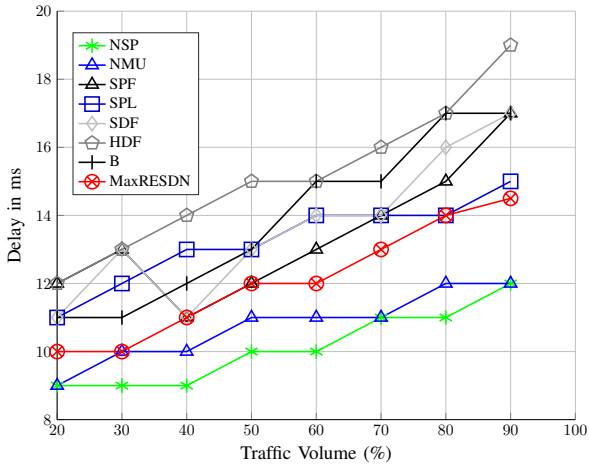
\begin{figure}[htb!]
		\begin{minipage}[b]{.70\textwidth}
		\scalebox{0.60}{
%
%
\definecolor{mycolor1}{rgb}{0,1,0}%
\definecolor{mycolor2}{rgb}{0,0,1}
\definecolor{mycolor3}{rgb}{0,0,0}%
\definecolor{mycolor4}{rgb}{0,0,0.8039}%
\definecolor{mycolor5}{rgb}{0.75294,0.75294,0.75294}%
\definecolor{mycolor6}{rgb}{0.50196,0.50196,0.50196}%
\definecolor{mycolor7}{rgb}{0,0,0}											
\definecolor{mycolor8}{rgb}{1,0,0}															
\begin{tikzpicture}

\begin{axis}[%
width=4.521in,
height=3.566in,
at={(1.322in,0.742in)},
scale only axis,
xmin=20,
xmax=100,
ymin=8,
ymax=20,
ylabel style={font=\color{white!15!black}},
ylabel={Delay in ms},
xlabel style={font=\large},
xlabel={Traffic Volume (\%)},
ylabel style={font=\large},
axis background/.style={fill=white},
axis x line*=bottom,
axis y line*=left,
xmajorgrids,
ymajorgrids,
legend style={at={(0.03,0.55)}, anchor=south west, legend columns=1, legend cell align=left, align=left, draw=white!15!black}
]
\addplot [ color=mycolor1, line width=1pt, mark size=4.5pt, mark=asterisk, mark options={solid, mycolor1}]
  table[row sep=crcr]{%
20	9\\
30	9\\
40	9\\
50	10\\
60	10\\
70	11\\
80	11\\
90	12\\
};
\addlegendentry{NSP}
\addplot [ color=mycolor2, line width=1pt, mark size=4.5pt, mark=triangle, mark options={solid, mycolor2}]
  table[row sep=crcr]{%
20	9\\
30	10\\
40	10\\
50	11\\
60	11\\
70	11\\
80	12\\
90	12\\
};
\addlegendentry{NMU}
\addplot [ color=mycolor3, line width=1pt, mark size=4.5pt, mark=triangle, mark options={solid, mycolor3}]
  table[row sep=crcr]{%
20	12\\
30	13\\
40	11\\
50	12\\
60	13\\
70	14\\
80	15\\
90	17\\
};
\addlegendentry{SPF}
\addplot [ color=mycolor4, line width=1pt, mark size=4.5pt, mark=square, mark options={solid, mycolor4}]
table[row sep=crcr]{%
20	11\\
30	12\\
40	13\\
50	13\\
60	14\\
70	14\\
80	14\\
90	15\\
};
\addlegendentry{SPL}
\addplot [ color=mycolor5, line width=1pt, mark size=4.5pt, mark=diamond, mark options={solid, mycolor5}]
table[row sep=crcr]{%
20	11\\
30	13\\
40	11\\
50	13\\
60	14\\
70	14\\
80	16\\
90	17\\
};
\addlegendentry{SDF}
\addplot [ color=mycolor6, line width=1pt, mark size=4.5pt, mark=pentagon, mark options={solid, mycolor6}]
  table[row sep=crcr]{%
20	12\\
30	13\\
40	14\\
50	15\\
60	15\\
70	16\\
80	17\\
90	19\\
};
\addlegendentry{HDF}

\addplot [ color=mycolor7, line width=1pt, mark size=4.5pt, mark=|, mark options={solid, mycolor7}]
  table[row sep=crcr]{%
20	11\\
30	11\\
40	12\\
50	13\\
60	15\\
70	15\\
80	17\\
90	17\\
};
\addlegendentry{B}

\addplot [ color=mycolor8, line width=1pt, mark size=4.5pt, mark=otimes, mark options={solid, mycolor8}]
  table[row sep=crcr]{%
20	10\\
30	10\\
40	11\\
50	12\\
60	12\\
70	13\\
80	14\\
90	14.5\\
};
\addlegendentry{MaxRESDN}

\end{axis}
\end{tikzpicture}%
}		
	\end{minipage}\qquad
	\hspace{-1cm}	
	\caption{Delay of heuristic algorithms in milliseconds  versus a range of traffic volume from 20\% to 90\%}
	\label{fig:delay} 
\end{figure}

Figure \ref{fig:delay} shows the delay of the heuristics algorithms measured in milliseconds with respect to traffic volume ranging from 20\% to 90\%. The NSP algorithm exhibits the least delay followed by NMU. The reason for this is because both NSP and NMU are initialized by the shortest paths. The SDF algorithm is 3 to 5 milliseconds worse in delay than the NSP algorithm. The MaxRESDN algorithm has a 1 millisecond to 3 milliseconds close to the NSP algorithm. It can be observed that for all of the energy saving heuristics, the delay tends to slightly increase with the increasing traffic volume. This is because all the energy saving algorithms attempt to minimize the number of links and switches used.

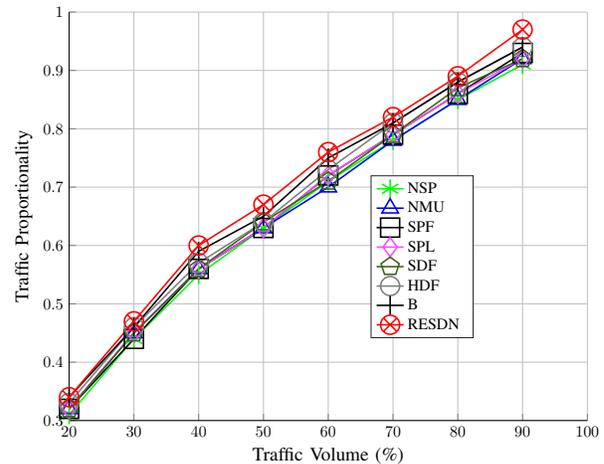
\begin{figure}[htb!]
		\begin{minipage}[b]{.70\textwidth}
		\scalebox{0.60}{
%
\definecolor{mycolor1}{rgb}{0,1,0}%
\definecolor{mycolor2}{rgb}{0,0,1}%
\definecolor{mycolor3}{rgb}{0,0,0}%
\definecolor{mycolor4}{rgb}{1,0.3,1}%
\definecolor{mycolor5}{rgb}{0.24,0.394,0.10}%
\definecolor{mycolor6}{rgb}{0.50196,0.50196,0.50196}%
\definecolor{mycolor7}{rgb}{0,0,0}	
\definecolor{mycolor8}{rgb}{1,0,0}	
\begin{tikzpicture}	
\begin{axis}[%
width=4.521in,	
height=3.566in,	
at={(1.322in,0.742in)},	
scale only axis,	
xmin=20,	
xmax=100,	
ymin=0.3,	
ymax=1,	
ylabel style={font=\color{white!15!black}},	
ylabel={Traffic Proportionality},	
xlabel style={font=\large},	
xlabel={Traffic Volume (\%)},	
ylabel style={font=\large},	
axis background/.style={fill=white},	
axis x line*=bottom,	
axis y line*=left,	
xmajorgrids,	
ymajorgrids,	
legend style={at={(0.582,0.202)}, anchor=south west, legend columns=1, legend cell align=left, align=left, draw=white!15!black}	
]	
\addplot [ color=mycolor1, line width=1.0pt, mark size=6pt, mark=asterisk, mark options={solid, mycolor1}]	
table[row sep=crcr]{%
20	0.31\\
30	0.44\\
40	0.55\\
50	0.63\\
60	0.71\\
70	0.78\\
80	0.85\\
90	0.91\\
};	
\addlegendentry{NSP}	
\addplot [ color=mycolor2, line width=1.0pt, mark size=6pt, mark=triangle, mark options={solid, mycolor2}]	
table[row sep=crcr]{%
20	0.32\\
30	0.45\\
40	0.56\\
50	0.63\\
60	0.7\\
70	0.78\\
80	0.85\\
90	0.92\\
};	
\addlegendentry{NMU}	
\addplot [ color=mycolor3, line width=1.0pt, mark size=6pt, mark=square, mark options={solid, mycolor3}]	
table[row sep=crcr]{%
20	0.32\\
30	0.44\\
40	0.56\\
50	0.63\\
60	0.72\\
70	0.79\\
80	0.86\\
90	0.93\\
};	
\addlegendentry{SPF}	
\addplot [ color=mycolor4, line width=1.0pt, mark size=6pt, mark=diamond, mark options={solid, mycolor4}]	
table[row sep=crcr]{%
20	0.32\\
30	0.45\\
40	0.56\\
50	0.63\\
60	0.72\\
70	0.79\\
80	0.86\\
90	0.92\\
};	
\addlegendentry{SPL}	
\addplot [ color=mycolor5, line width=1.0pt, mark size=6pt, mark=pentagon, mark options={solid, mycolor5}]	
table[row sep=crcr]{%
20	0.32\\
30	0.45\\
40	0.56\\
50	0.64\\
60	0.71\\
70	0.79\\
80	0.87\\
90	0.92\\
};	
\addlegendentry{SDF}	
\addplot [ color=mycolor6, line width=1.0pt, mark size=6pt, mark=o, mark options={solid, mycolor6}]	
table[row sep=crcr]{%
20	0.33\\
30	0.46\\
40	0.57\\
50	0.64\\
60	0.73\\
70	0.81\\
80	0.88\\
90	0.94\\
};	
\addlegendentry{HDF}	
\addplot [ color=mycolor7, line width=1.0pt, mark size=6pt, mark=|, mark options={solid, mycolor7}]	
table[row sep=crcr]{%
20	0.34\\
30	0.46\\
40	0.59\\
50	0.65\\
60	0.75\\
70	0.81\\
80	0.88\\
90	0.94\\
};	
\addlegendentry{B}	
\addplot [ color=mycolor8, line width=1.0pt, mark size=6pt, mark=otimes, mark options={solid, mycolor8}]	
table[row sep=crcr]{%
20	0.34\\
30	0.47\\
40	0.6\\
50	0.67\\
60	0.76\\
70	0.82\\
80	0.89\\
90	0.97\\
};	
\addlegendentry{RESDN }	
\end{axis}	
\end{tikzpicture}%
}		
	\end{minipage}\qquad
	\hspace{-1cm}	
	\caption{Traffic proportionality for the range of traffic volume from 20\% to 90\%}
	\label{fig:propotionality} 
\end{figure}

Figure \ref{fig:propotionality} shows the traffic proportionality of heuristics for the range of traffic volume from 20\% to 90\%. The MaxRESDN heuristics have demonstrated the maximum traffic proportionality value. The B heuristics, which is an attempt of keeping the trade-off between performance and energy saving, has a closer traffic proportionality to our approach. 

\begin{figure}[htb!]
	\begin{minipage}[b]{.70\textwidth}
		\scalebox{0.60}{
%
%
\definecolor{mycolor1}{rgb}{0.00000,0.44700,0.74100}%
\definecolor{mycolor2}{rgb}{0.85000,0.32500,0.09800}%
\definecolor{mycolor3}{rgb}{0.83137,0.81569,0.78431}%
\begin{tikzpicture}

\begin{axis}[%
width=4.521in,
height=3.566in,
at={(2.6in,1.095in)},
scale only axis,
bar shift auto,
log origin=infty,
xmin=0.52,
xmax=8.5,
xtick={1,2,3,4,5,6,7,8},
xticklabels={{\large{NSP}},\large{{NMU}},\large{{SPF}},\large{{SPL}},\large{{SDF}},\large{{HDF}},\large{{B}},\large{{MaxRESDN}}},
xlabel style={font=\color{white!15!black}},
xlabel={Heuristics},
xlabel style={font=\large},
ymin=0.6,
ymax=0.70,
ytick={0.6,0.62,0.64,0.66,0.68,0.70},
yticklabels={\large{0.6},\large{0.62},\large{0.64},\large{0.66},\large{0.68},\large{0.70}},
ylabel style={font=\color{white!15!black}},
ylabel={Average Traffic Proportionality},
ylabel style={font=\large},
axis background/.style={fill=white},
axis x line*=bottom,
axis y line*=left,
xmajorgrids,
ymajorgrids,
legend style={at={(0.133,0.681)}, anchor=south west, legend cell align=left, align=left, draw=white!15!black}
]

\addplot[ybar, bar width=0.478, fill=mycolor3, draw=black, area legend] table[row sep=crcr] {%
1	0.65\\
2	0.66\\
3	0.66\\
4	0.66\\
5	0.66\\
6	0.67\\
7	0.68\\
8	0.7\\
};

\end{axis}
\end{tikzpicture}%
}		
	\end{minipage}\qquad
	\hspace{-1cm}	
	\caption{Average traffic proportionality of MaxRESDN as compared to other heuristics}
	\label{fig:avgpropotionality} 
\end{figure}
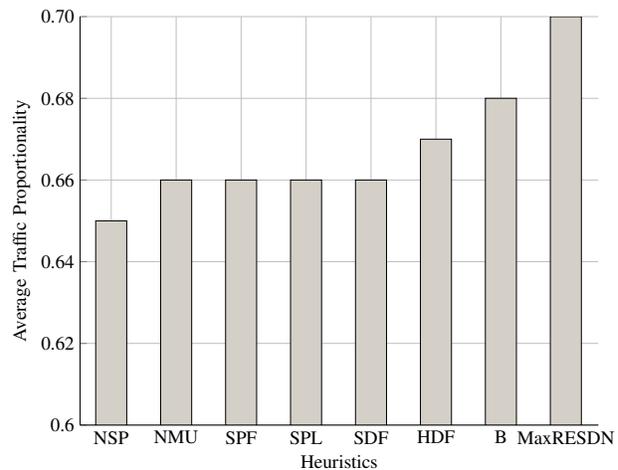

Figure \ref{fig:avgpropotionality} shows average traffic proportionality of all traffic volumes from 20\% to 90\% for MaxRESDN algorithm compared with the other algorithms (in Table \ref{tbl:algorithms}). The MaxRESDN heuristics exhibits the largest traffic proportionality in terms of link energy consumption. NSP algorithm has the lowest traffic proportionality. As compared to those heuristics which give priority to network performance such as NSP and SPF, our approach is 4 to 5\% better in traffic proportionality, and 3 to 4\% better in traffic proportionality than heuristics that give priority to energy saving and maximizing utility such as NSP and HDF.

\subsection{Analysis of Utility Parameters}\label{subsec:anlys}

A challenge with the MaxRESDN algorithm is its dependence on the link utility interval parameters. The trends of link utility interval parameters $Umin$ and $Umax$ versus traffic volume and percentage of links saved are analyzed in this subsection. For the fixed $Umax$ value to 95\%, we study the effect of the $Umin$ parameter on the percentage of links saved for traffic volume ranging from 20\% to 90\% $Umin$. Likewise, for the fixed $Umin$ value to 30\%, we study the effect of $Umax$ value on the percentage of links saved. Understanding the relations and the trend for values of $Umin$ and $Umax$  versus the traffic volume would be significant in predicting the utility parameters for future use.  

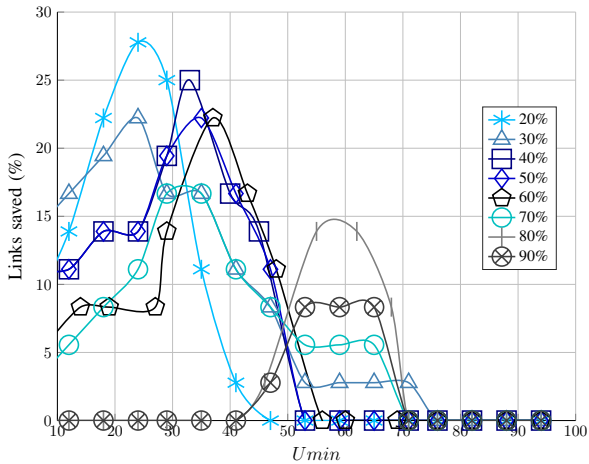
\begin{figure}[htb!]
	\begin{minipage}[b]{.70\textwidth}
		\scalebox{0.60}{
%
%
\definecolor{mycolor1}{rgb}{0,0.7490,1}%
\definecolor{mycolor2}{rgb}{0.2745,0.5098,0.70588}%
\definecolor{mycolor3}{rgb}{0,0,0.5019}%
\definecolor{mycolor4}{rgb}{0,0,0.8039}%
\definecolor{mycolor5}{rgb}{0.0,0.0,0.0}%
\definecolor{mycolor6}{rgb}{0.0.75,0.750,0.75}%
\definecolor{mycolor7}{rgb}{0.50,0.50,0.50}
\definecolor{mycolor8}{rgb}{0.25,0.25,0.25}
\begin{tikzpicture}

\begin{axis}[%
width=4.521in,
height=3.566in,
at={(1.322in,0.742in)},
scale only axis,
xmin=10,
xmax=100,
ymin=0,
ymax=30,
ylabel style={font=\color{white!15!black}},
ylabel={Links saved (\%)},
xlabel style={font=\large},
xlabel={$Umin$},
ylabel style={font=\large},
axis background/.style={fill=white},
axis x line*=bottom,
axis y line*=left,
xmajorgrids,
ymajorgrids,
legend style={at={(0.82,0.369)}, anchor=south west, legend columns=1, legend cell align=left, align=left, draw=white!15!black}
]
\addplot [smooth, color=mycolor1, line width=1.0pt, mark size=6pt, mark=asterisk, mark options={solid, mycolor1}]
  table[row sep=crcr]{%
6	8.33\\
12	13.89\\
18	22.22\\
24	27.78\\
29	25\\
35	11.11\\
41	2.78\\
47	0\\
53	0\\
59	0\\
65	0\\
71	0\\
76	0\\
82	0\\
88	0\\
94	0\\
};
\addlegendentry{20\%}

\addplot [smooth, color=mycolor2, line width=1.0pt, mark size=6pt, mark=triangle, mark options={solid, mycolor2}]
  table[row sep=crcr]{%
6	13.89\\
12	16.67\\
18	19.44\\
24	22.22\\
29	16.67\\
35	16.67\\
41	11.11\\
47	8.33\\
53	2.78\\
59	2.78\\
65	2.78\\
71	2.78\\
76	0\\
82	0\\
88	0\\
94	0\\
};
\addlegendentry{30\%}

\addplot [smooth, color=mycolor3, line width=1.0pt, mark size=6pt, mark=square, mark options={solid, mycolor3}]
  table[row sep=crcr]{%
6	11.11\\
12	11.11\\
18	13.89\\
24	13.89\\
29	19.44\\
33	25\\
40	16.67\\
45	13.89\\
53	0\\
59	0\\
65	0\\
71	0\\
76	0\\
82	0\\
88	0\\
94	0\\
};
\addlegendentry{40\%}
\addplot [smooth, color=mycolor4, line width=1.0pt, mark size=6pt, mark=diamond, mark options={solid, mycolor4}]
table[row sep=crcr]{%
	6	11.11\\
	12	11.11\\
	18	13.89\\
	24	13.89\\
	29	19.44\\
	35	22.22\\
	41	16.67\\
	47	11.11\\
	53	0\\
	59	0\\
	65	0\\
	71	0\\
	76	0\\
	82	0\\
	88	0\\
	94	0\\
};
\addlegendentry{50\%}
\addplot [smooth, color=mycolor5, line width=1.0pt, mark size=6pt, mark=pentagon, mark options={solid, mycolor5}]
table[row sep=crcr]{%
	8	5.56\\
	14	8.33\\
	19	8.33\\
	27	8.33\\
	29	13.89\\
	37	22.22\\
	43	16.67\\
	48	11.11\\
	56	0\\
	60	0\\
	69	0\\
	71	0\\
	76	0\\
	82	0\\
	88	0\\
	94	0\\
};
\addlegendentry{60\%}

\addplot [smooth, color=mycolor6, line width=1.0pt, mark size=6pt, mark=o, mark options={solid, mycolor6}]
  table[row sep=crcr]{%
6	2.78\\
12	5.56\\
18	8.33\\
24	11.11\\
29	16.67\\
35	16.67\\
41	11.11\\
47	8.33\\
53	5.56\\
59	5.56\\
65	5.56\\
71	0\\
76	0\\
82	0\\
88	0\\
94	0\\
};
\addlegendentry{70\%}

\addplot [smooth, color=mycolor7, line width=1.0pt, mark size=6pt, mark=|, mark options={solid, mycolor7}]
  table[row sep=crcr]{%
6	0\\
12	0\\
18	0\\
24	0\\
29	0\\
35	0\\
41	0\\
46	2.78\\
55	13.89\\
62	13.89\\
68	8.33\\
71	0\\
76	0\\
82	0\\
88	0\\
94	0\\
};
\addlegendentry{80\%}

\addplot [smooth, color=mycolor8, line width=1.0pt, mark size=6pt, mark=otimes, mark options={solid, mycolor8}]
  table[row sep=crcr]{%
6	0\\
12	0\\
18	0\\
24	0\\
29	0\\
35	0\\
41	0\\
47	2.78\\
53	8.33\\
59	8.33\\
65	8.33\\
71	0\\
76	0\\
82	0\\
88	0\\
94	0\\
};
\addlegendentry{90\%}

\end{axis}
\end{tikzpicture}%
}		
	\end{minipage}\qquad
	\caption{Effect of utility parameter $Umin$ values on the percentage of links saved for different traffic volumes ranging from 20\% to 90\% by fixing $Umax$ to 95\%}
	\label{fig:umin} 
\end{figure}

Figure \ref{fig:umin} shows the effect of the $Umin$ parameter on the percentage of links saved for traffic volumes 20\% to 90\%. Each line represents a traffic volume. Fixing $Umax$ to 95\%, and ranging $Umin$ value from 10\% to 90\%, the trend shows that the value of the $Umin$ that maximizes the links saved increases till it reaches a peak then drops. As the traffic volume increases the peak $Umin$ value increases. The analysis exhibits similar trend for different traffic volumes. 

 \begin{figure}[htb!]
	\begin{minipage}[b]{.70\textwidth}
		\scalebox{0.60}{
%
%
\definecolor{mycolor1}{rgb}{0,0.7490,1}%
\definecolor{mycolor2}{rgb}{0.2745,0.5098,0.70588}%
\definecolor{mycolor3}{rgb}{0,0,0.5019}%
\definecolor{mycolor4}{rgb}{0,0,0.8039}%
\definecolor{mycolor5}{rgb}{0.0,0.0,0.0}%
\definecolor{mycolor6}{rgb}{0.0.75,0.750,0.75}%
\definecolor{mycolor7}{rgb}{0.50,0.50,0.50}
\definecolor{mycolor8}{rgb}{0.25,0.25,0.25}
\begin{tikzpicture}

\begin{axis}[%
width=4.521in,
height=3.566in,
at={(1.322in,0.742in)},
scale only axis,
xmin=40,
xmax=100,
ymin=0,
xmajorgrids,
ymax=30,
xlabel style={font=\large},
ylabel style={font=\large},
ylabel={Links Saved (\%)},
ymajorgrids,
xlabel={$Umax$},
axis background/.style={fill=white},
axis x line*=bottom,
axis y line*=left,
legend style={at={(0.02,0.48)}, anchor=south west, legend columns=1, legend cell align=left, align=left, draw=white!15!black}
]
\addplot [smooth, color=mycolor1, line width=1.0pt, mark size=6pt, mark=asterisk, mark options={solid, mycolor1}]
  table[row sep=crcr]{%
30	8.33\\
40	11.11\\
50	13.88\\
60	16.66\\
70	16.66\\
80	19.44\\
90	18.22\\
100	18.22\\
};
\addlegendentry{20\%}

\addplot [smooth, color=mycolor2, line width=1.0pt, mark size=6pt, mark=triangle, mark options={solid, mycolor2}]
  table[row sep=crcr]{%
30	0\\
40	0\\
50	2.77\\
60	5.55\\
70	8.33\\
80	13.88\\
90	22.22\\
100	22.22\\
};
\addlegendentry{30\%}

\addplot [smooth, color=mycolor3, line width=1.0pt, mark size=6pt, mark=square, mark options={solid, mycolor3}]
  table[row sep=crcr]{%
30	0\\
40	0\\
50	0\\
60	0\\
70	5.55\\
80	11.11\\
90	22.22\\
100	22.22\\
};
\addlegendentry{40\%}
\addplot [smooth, color=mycolor4, line width=1.0pt, mark size=6pt, mark=diamond, mark options={solid, mycolor4}]
table[row sep=crcr]{%
	30	0\\
	40	0\\
	50	0\\
	60	0\\
	70	2.77\\
	80	13.88\\
	90	27.77\\
	100	27.77\\
};
\addlegendentry{50\%}

\addplot [smooth, color=mycolor5, line width=1.0pt, mark size=6pt, mark=pentagon, mark options={solid, mycolor5}]
table[row sep=crcr]{%
	30	0\\
	40	0\\
	50	0\\
	60	0\\
	70	0\\
	80	5.55\\
	90	22.22\\
	100	22.22\\
};
\addlegendentry{60\%}

\addplot [smooth, color=mycolor6, line width=1.0pt, mark size=6pt, mark=o, mark options={solid, mycolor6}]
  table[row sep=crcr]{%
30	0\\
40	0\\
50	0\\
60	0\\
70	0\\
80	8.33\\
90	19.44\\
100	19.44\\
};
\addlegendentry{70\%}

\addplot [smooth, color=mycolor7, line width=1.0pt, mark size=6pt, mark=|, mark options={solid, mycolor7}]
  table[row sep=crcr]{%
30	0\\
40	0\\
50	0\\
60	0\\
70	5.55\\
80	13.88\\
90	19.44\\
100	19.44\\
};
\addlegendentry{80\%}

\addplot [smooth, color=mycolor8, line width=1.0pt, mark size=6pt, mark=otimes, mark options={solid, mycolor8}]
  table[row sep=crcr]{%
30	0\\
40	0\\
50	0\\
60	2.77\\
70	2.77\\
80	8.33\\
90	13.88\\
100	13.88\\
};
\addlegendentry{90\%}

\end{axis}
\end{tikzpicture}%
}		
	\end{minipage}\qquad
	\hspace{-1cm}	
	
	\caption{Effect of utility parameter $Umax$ values on the percentage of links saved for different traffic volumes ranging from 20\% to 90\% by fixing $Umin$ to 30\%}
	\label{fig:umax}	
\end{figure}
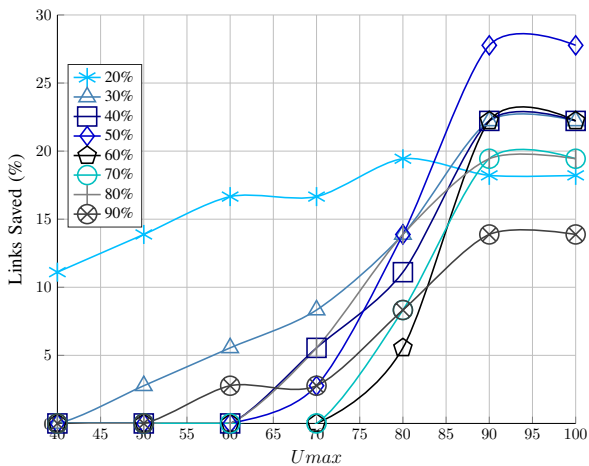

Figure \ref{fig:umax} shows the effect of the $Umax$ value on the percentage of links saved for traffic volumes from 20\% to 90\%. Each line shows the traffic volume the experiment is conducted. The $Umin$ parameter is set to 30\% and the $Umax$ value ranges from 40\% to 100\%. For every flow size, the $Umax$ value that maximizes the number of links saved increases till it reaches its peak then it mainly remains constant except for the 20\% traffic volume where there is a slight decrease. 

\begin{figure}[htb!]
	
	\begin{minipage}[b]{.70\textwidth}
		\scalebox{0.60}{
%
%
\definecolor{mycolor1}{rgb}{0,0.7490,1}%
\definecolor{mycolor2}{rgb}{0.2745,0.5098,0.70588}%
\definecolor{mycolor3}{rgb}{0,0,0.5019}%
\definecolor{mycolor4}{rgb}{0,0,0.8039}%
\definecolor{mycolor5}{rgb}{0.0,0.0,0.0}%
\definecolor{mycolor6}{rgb}{0.0.75,0.750,0.75}%
\definecolor{mycolor7}{rgb}{0.50,0.50,0.50}
\definecolor{mycolor8}{rgb}{0.25,0.25,0.25}
\begin{tikzpicture}

\begin{axis}[%
width=4.521in,
height=3.566in,
at={(1.322in,0.742in)},
scale only axis,
xmin=20,
xmax=100,
xtick={20,30,40,50,60,70,80,90,100},
xlabel style={font=\color{white!15!black}},
xlabel={Traffic Volume (\%)},
ylabel style={font=\large},
ymin=20,
ymax=60,
ytick={20,30,40,50,60},
xlabel style={font=\large},
ylabel={$Umin$},
xlabel style={font=\large},
ymajorgrids,
xmajorgrids,
axis background/.style={fill=white},
axis x line*=bottom,
axis y line*=left,
legend style={at={(0.12,0.469)}, anchor=south west, legend columns=1, legend cell align=left, align=left, draw=white!15!black}
]
\addplot [smooth, color=blue, line width=1.0pt, mark size=6pt, mark=asterisk, mark options={solid, blue}]
  table[row sep=crcr]{%
20	19\\
30	20\\
40	24\\
50	30\\
60	31\\
70	38\\
80	45\\
90	46\\
};
\addlegendentry{$Umax$=85\%}

\addplot [smooth, color=black, line width=1.0pt, mark size=6pt, mark=triangle, mark options={solid, black}]
  table[row sep=crcr]{%
20	22\\
30	23\\
40	31\\
50	33\\
60	44\\
70	47\\
80	48\\
90	48\\
};
\addlegendentry{$Umax$=90\%}
\addplot [smooth, color=green, line width=1.0pt, mark size=6pt, mark=square, mark options={solid, green}]
  table[row sep=crcr]{%
20	24\\
30	25\\
40	33\\
50	35\\
60	37\\
70	46\\
80	55\\
90	56\\
};
\addlegendentry{$Umax$=95\%}
\end{axis}
\end{tikzpicture}%
}
	\end{minipage}\qquad
	\caption{The effect of the traffic volume ranging from 20\% to 90\% on the value of $Umin$  and $Umax$ that maximizes the percentage of links saved  fixing $Umax$ values to 85\%, 90\% and 95\%} 
	\label{fig:tanalysisofumin}
\end{figure}
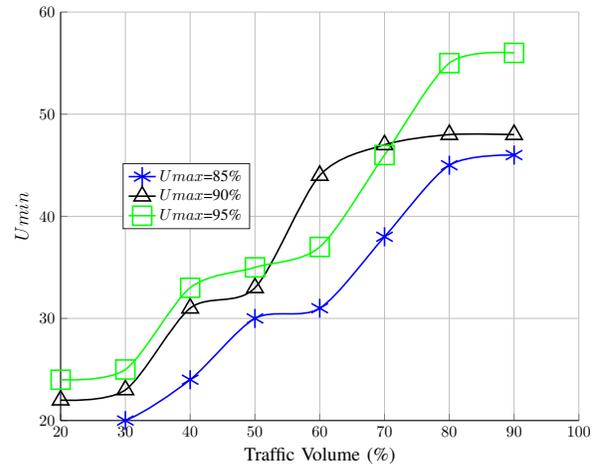

 Figure \ref{fig:tanalysisofumin} shows the relationship between the traffic volume and the $Umin$ value that maximizes the energy saving. It shows that as the traffic volume increases the $Umin$ that maximizes energy saving increases. The reason for this is because the utilities of the links increase with traffic volume. For example when the percentage of traffic is 90\%, the link with minimum utility is above 40\%. Which means that if we pick $Umin$ value less than 40\%, the candidate list would be empty. Similarly as the traffic volume decrease to 20\%, the link utilities starts from 12\% and we will have more links in the candidate list. According to results on figures \ref{fig:umin} and \ref{fig:tanalysisofumin}, the problem that need to be solved hence is to find or estimate the peak $Umin$ value for a given traffic volume.

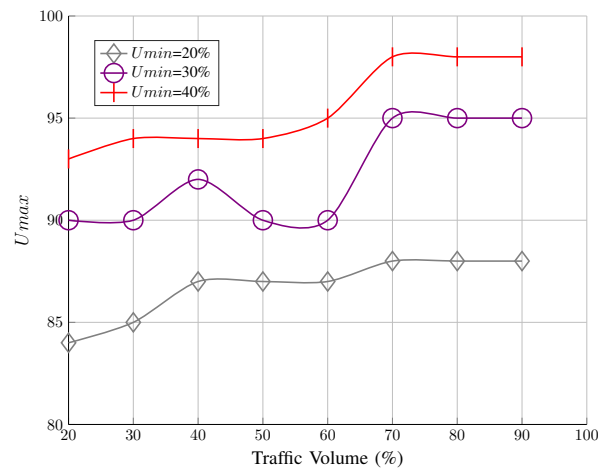
\begin{figure}[htb!]
	\begin{minipage}[b]{.70\textwidth}
		\scalebox{0.60}{
%
%
\definecolor{mycolor1}{rgb}{0,0.7490,1}%
\definecolor{mycolor2}{rgb}{0.2745,0.5098,0.70588}%
\definecolor{mycolor3}{rgb}{0,0,0.5019}%
\definecolor{mycolor4}{rgb}{0,0,0.8039}%
\definecolor{mycolor5}{rgb}{0.0,0.0,0.0}%
\definecolor{mycolor6}{rgb}{0.0.75,0.750,0.75}%
\definecolor{mycolor7}{rgb}{0.50,0.50,0.50}
\definecolor{mycolor8}{rgb}{0.25,0.25,0.25}
\begin{tikzpicture}

\begin{axis}[%
width=4.521in,
height=3.566in,
at={(1.322in,0.742in)},
scale only axis,
xmin=20,
xmax=100,
xtick={20,30,40,50,60,70,80,90,100},
xlabel style={font=\color{white!15!black}},
xlabel={Traffic Volume (\%)},
xmajorgrids,
ymin=80,
ymax=100,
ytick={80,85,90,95,100},
ylabel style={font=\color{white!15!black}},
ylabel={$Umax$},
ymajorgrids,
axis background/.style={fill=white},
xlabel style={font=\large},
ylabel style={font=\large},
axis x line*=bottom,
axis y line*=left,
legend style={at={(0.05,0.78)}, anchor=south west, legend columns=1, legend cell align=left, align=left, draw=white!15!black}
]
\addplot [smooth, color=gray, line width=1.0pt, mark size=6pt, mark=diamond, mark options={solid, gray}]
  table[row sep=crcr]{%
20	84\\
30	85\\
40	87\\
50	87\\
60	87\\
70	88\\
80	88\\
90	88\\
};
\addlegendentry{$Umin$=20\%}

\addplot [smooth, color=violet, line width=1.0pt, mark size=6pt, mark=o, mark options={solid, violet}]
  table[row sep=crcr]{%
20	90\\
30	90\\
40	92\\
50	90\\
60	90\\
70	95\\
80	95\\
90	95\\
};
\addlegendentry{$Umin$=30\%}
\addplot [smooth, color=red, line width=1.0pt, mark size=6pt, mark=|, mark options={solid, red}]
  table[row sep=crcr]{%
20	93\\
30	94\\
40	94\\
50	94\\
60	95\\
70	98\\
80	98\\
90	98\\
};
\addlegendentry{$Umin$=40\%}
\end{axis}
\end{tikzpicture}%
}
	\end{minipage}\qquad
	\caption{The effect of the traffic volume ranging from 20\% to 90\% on the value of  $Umax$ that maximizes the percentage of links saved fixing $Umin$ values to 20\%, 30\%, and 40\%} 
	 \label{fig:tanalysisofumax}	
\end{figure}

Figure \ref{fig:tanalysisofumax} illustrates the trend of $Umax$ value  in relation to traffic volume by fixing the value of $Umin$ the at 30\%. It demonstrates that as the volume of traffic increases, the $Umax$ maximizing the energy saving exhibits an increasing trend. It also shows that for a traffic volume which is more than 72\%, the $Umax$ value keep a constant value near 95\%. According to the results depicted on Figure \ref{fig:tanalysisofumax} and ref{fig:tanalysisofumin}, picking the right value of $Umin$ and $Umax$ has a direct impact on the performance of the RESDN algorithm.   

\section{Conclusions} \label{sec:conclusion}

The flexibility of network control by SDN can be used for energy efficient routing. However, minimizing energy consuming devices without compromising performance is a challenging task. We proposed energy efficiency metric named RESDN (Ratio for Energy Saving in SDN) that puts the utility of links under consideration. Unlike other energy efficiency metrics, RESDN is based on the link utility interval parameters, and captures the dynamic network changes in SDN. We also provided an IP formulation with the objective of increasing the RESDN of a network environment and a heuristics named MaxRESDN that maximizes energy saving. The experiments are conducted on Mininet network emulator and POX controller using the GEANT network topology and dynamic traffic traces. We also simulated the power consumption of OvSwitch, NEC PF5240, and Zodiac FX OpenFlow switches.

Experimental results show that maximizing the RESDN value improves energy efficiency while maintaining acceptable network performance. MaxRESDN is up to 30\%  better in the number of links and achieves up to 14.7 watts, 10 watts, and 3.2 watts less power consumption for NEC, OVS and Zodiac FX switches respectively as compared to other utility-based heuristics for energy efficient routing. The maximum RESDN value is achieved by our proposed MaxRESDN method, which also performs close to the solutions that give priority to network performance in terms of average path length, throughput, and delay. Our approach not only does have 3 to 5\% higher traffic proportionality but also maintains the trade-off between network performance and energy efficiency. Since the performance of the MaxRESDN heuristics greatly depends on the value of utility interval parameters, we performed a detailed analysis of the parameters with regard to traffic volume. It is demonstrated that the values of utility interval parameters directly affect the efficiency of the algorithm. 

As future work, we aim to utilize supervised and reinforcement machine learning techniques to predict the link utility interval parameters in an efficient manner. We also plan to apply the idea behind RESDN metric to incorporate end systems in a data center platform, and improve the scalability through parallelizing the MaxRESDN algorithm to handle multiple flow arrivals concurrently. 

\section*{Acknowledgments}
This work was partially supported by the COST (European Cooperation in Science and Technology) framework, under Action IC0804 (Energy Efficiency in Large Scale Distributed Systems), and by TUBITAK (The Scientific and Technical Research Council of Turkey) under Grant 109M761.
%

\bibliographystyle{IEEEtran} 
\bibliography{RESDNreference}

\vskip -1\baselineskip plus -1fil
\begin{IEEEbiography}
    [{\includegraphics[width=1in,height=1.25in,clip,keepaspectratio]{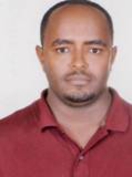}}]
     {Beakal G. Assefa} is a Ph.D. Candidate, teaching and research assistant in the Department of Computer Engineering at Koç University. He received his M.S. in Computer Engineering from Izmir Institute of Technology in 2012. He received his B.S. degree in Computer Science and Information Technology from Haramaya University, Ethiopia. His research interests include databases, software-defined networks,energy efficiency, machine learning, and distributed systems
\end{IEEEbiography}

\vskip -1\baselineskip plus -1fil

\begin{IEEEbiography}
    [{\includegraphics[width=1in,height=1.25in,clip,keepaspectratio]{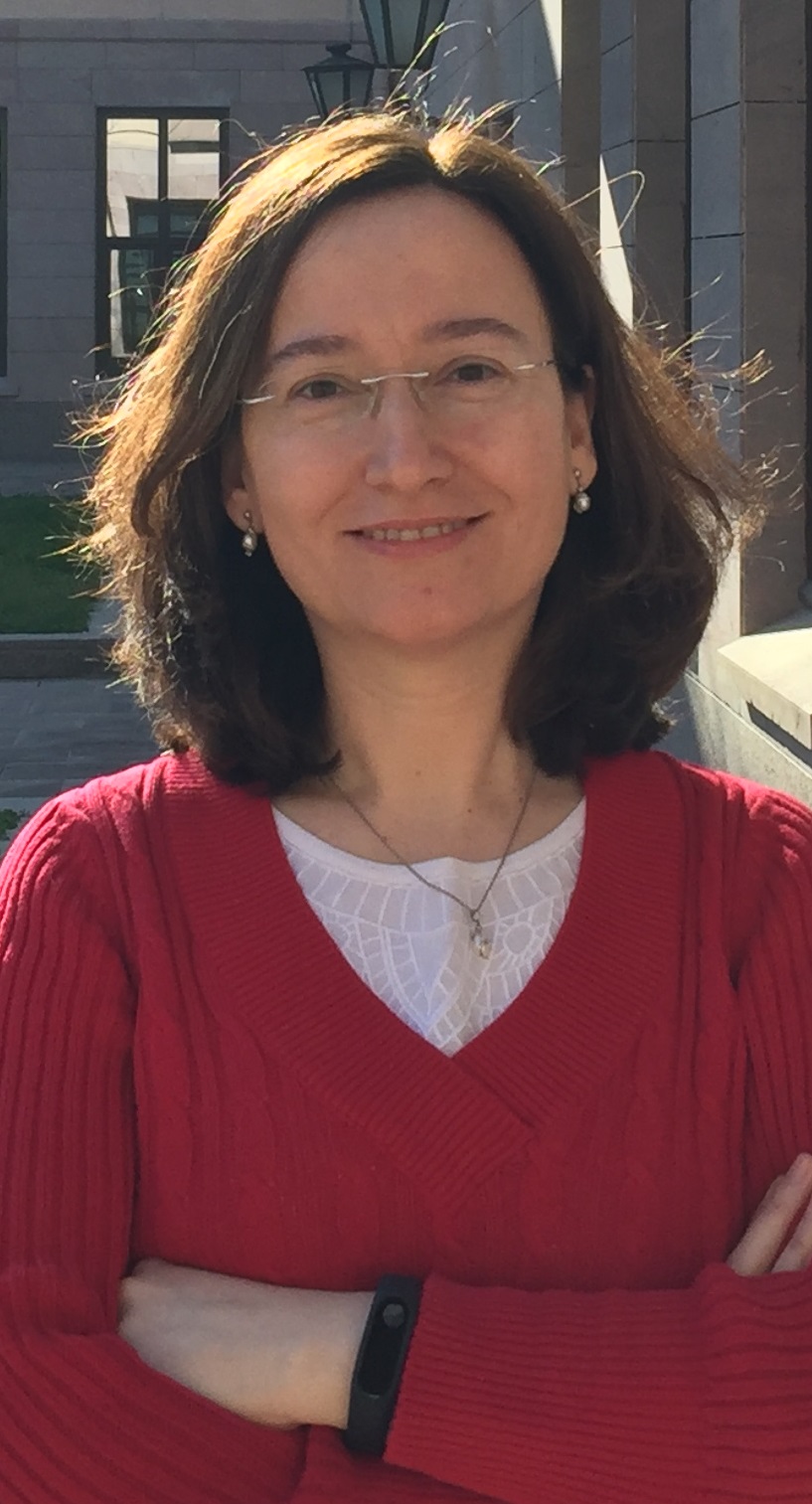}}]
    {Öznur Özkasap} received the Ph.D. degree in Computer Engineering from Ege University in 2000. From 1997 to 1999, she was a Graduate Research Assistant with the Department of Computer Science, Cornell University, where she completed her Ph.D. dissertation. She is currently a Professor with the Department of Computer Engineering, Koç University, which she joined in 2000. Her research interests include distributed systems, multicast protocols, peer-to-peer systems, bio-inspired distributed algorithms, mobile and vehicular ad hoc networks, energy efficiency, cloud computing, and computer networks. Prof. Özkasap is IEEE Senior Member, serves as an Area Editor of the Future Generation Computer Systems journal, Elsevier Science, and National Representative of ACM-W Europe. She is a recipient of the Turk Telekom Collaborative Research Awards, the Career Award of TUBITAK (The Scientific and Technological Research Council of Turkey), and TUBITAK/NATO A2 Ph.D. Research Scholarship Abroad, and she was awarded Teaching Innovation Grants by Koç University.

\end{IEEEbiography}

\end{document}